\documentclass[lettersize,journal]{IEEEtran}
\usepackage{amsmath,amsfonts}
\usepackage{algorithmic}
\usepackage{algorithm}
\usepackage{array}
\usepackage[caption=false,font=footnotesize,labelfont=rm,textfont=rm]{subfig}
\usepackage{textcomp}
\usepackage{stfloats}
\usepackage{url}
\usepackage{verbatim}
\usepackage{graphicx}
\usepackage{cite}
\usepackage{stfloats}
\usepackage{makecell}
\usepackage{multirow}
\usepackage{amsthm,amsmath,amssymb}
\usepackage{mathrsfs}
\usepackage{epsfig}
\usepackage{epstopdf}
\usepackage{diagbox}
\usepackage{soul}
\usepackage{ulem}
\usepackage{color, xcolor}
\begin{document}

\title{A Shared Cluster-based Stochastic Channel Model for Integrated Sensing and Communication Systems}

\author{Yameng Liu, \IEEEmembership{Graduate Student Member, IEEE}, Jianhua Zhang, \IEEEmembership{Senior Member, IEEE}, \\
Yuxiang Zhang, \IEEEmembership{Member, IEEE}, Zhiqiang Yuan, \IEEEmembership{Graduate Student Member, IEEE}, \\
and Guangyi Liu, \IEEEmembership{Member, IEEE}
\thanks{Copyright (c) 2015 IEEE. Personal use of this material is permitted. However, permission to use this material for any other purposes must be obtained from the IEEE by sending a request to pubs-permissions@ieee.org.}
\thanks{This work was supported in part by National Natural Science Foundation of China under Grant 92167202, in part by National Science Foundation for Distinguished Young Scholars under Grant 61925102, in part by National Natural Science Foundation of China under Grant 62201087, Grant 62101069, and Grant 62201086, and in part by the Beijing University of Posts and Telecommunications-China Mobile Research Institute Joint Innovation Center. \textit{(Corresponding author: Jianhua Zhang; Yuxiang Zhang.)}}
\thanks{Yameng Liu, Jianhua Zhang, Yuxiang Zhang, and Zhiqiang Yuan are with the State Key Laboratory of Networking and Switching Technology, Beijing University of Posts and Telecommunications, Beijing 100876, China (email: liuym@bupt.edu.cn; jhzhang@bupt.edu.cn; zhangyx@bupt.edu.cn; yuanzhiqiang@bupt.edu.cn).}
\thanks{Guangyi Liu is with China Mobile Research Institution, Beijing 100053, China, (email: liuguangyi@chinamobile.com).}}

\markboth{Journal of \LaTeX\ Class Files,~Vol.~14, No.~8, August~2021}%
{Shell \MakeLowercase{\textit{et al.}}: A Sample Article Using IEEEtran.cls for IEEE Journals}


\maketitle

\begin{abstract}
Integrated Sensing And Communication (ISAC) has been recognized as a promising technology in the 6G communication. A realistic channel model is a prerequisite for designing ISAC systems. Most existing channel models independently generate the communication and sensing channels under the same framework. However, due to the multiplexing of hardware resources and the same environment, signals enabled for communication and sensing may experience shared propagation scatterers. This practical sharing feature necessities the joint generation of communication and sensing channels for realistic modeling, where the shared clusters (contributed by the shared scatterers) should be reconstructed. In this paper, we first conduct communication and sensing channel measurements for an indoor scenario at 28 GHz. The power-angular-delay profiles of multipath components are obtained, and the shared scatterers by communication and sensing channels are intuitively observed. Then, a stochastic ISAC channel model is proposed to capture the sharing feature, where shared and non-shared clusters by the two channels are defined and superimposed. To extract those clusters from measured ISAC channels, a KPowerMeans-based joint clustering algorithm is novelly introduced. Finally, stochastic channel characteristics are analyzed, and empirical simulations validate that the channel Sharing Degree (SD) increases with more shared clusters. The proposed model can realistically capture the sharing feature of ISAC channels and is able to evaluate and simulate the channel SD values, which is valuable for the design and deployment of ISAC systems.
\end{abstract}


\section{Introduction}\label{section1}
\IEEEPARstart{N}{o}wadays, Integrated Sensing And Communication (ISAC) has been recognized as a promising technology to achieve ubiquitous sensing and digital twin for the sixth generation (6G) systems \cite{liu2020vision}. The communication purpose is to transmit information, while sensing aims at the detection and identification of the targets. Compared to conventional systems with separate devices, ISAC technology realizes the capability of integrating functions into one system, enabling the base stations or terminals to communicate while sensing the surrounding environment \cite{kumari2017ieee,nie2022predictive,zhang2018multibeam,wang2022empirical}. By sharing a majority of the software, hardware, and information resources, ISAC systems bring tremendous advantages to improving spectrum utilization and reducing costs \cite{liu2022survey,cui2021integrating,zhang2021overview,pucci2022system}. 

Realistic channel modeling is the prerequisite for the design and deployment of ISAC systems \cite{zhang2020channel,yuan2022spatial}. Communication and sensing channels are intrinsically different because of different propagation configurations, e.g., transmitter-receiver (TX-RX) relative positions. In the existing research, ISAC channel models are widely utilized in ISAC systems to evaluate upper-level technology development and system design \cite{nguyen2022access,rahman2019framework,zhang2022integrated,yuan2020spatio}, where communication and sensing channels are generated independently. However, due to the multiplexing of hardware resources (e.g., antennas) and the same environment, some objects might serve as shared propagation scatterers for both the communication and sensing channels. This sharing feature has been observed in several multi-link channel measurements, and the effects on correlation and capacity of the shared clusters (contributed by the shared scatterers) are studied \cite{poutanen2010significance,poutanen2011multi}. The sharing of communication and sensing channels is necessary to leverage channel information obtained by one system to enhance the performance of the other. Hence, unlike existing models that independently generate communication and sensing channels, it is imperative to carefully consider the observed sharing feature of scatterers and clusters in ISAC channel modeling.

The relationship between communication and sensing channels has attracted attention in a few existing ISAC research. In \cite{liu2020joint}, communication scatterers are assumed to be part of the sensing targets, and a dual-functional ISAC system framework at millimeter wave (mmWave) bands is proposed for unifying sensing and communication operations. In \cite{chen2021code}, the sensing path is simply assumed to be the round-trip of the communication path, with the delay and Doppler of sensing channels defined as twice that of the communication channels. Considering the sharing feature in ISAC systems, studies on the similarity of azimuth power spectrum of communication and sensing are performed using Ray-tracing simulations in \cite{ali2020passive,gonzalez2016radar}. Moreover, a measurement campaign for both communication and sensing channels is conducted in \cite{graff2019measuring} to investigate the realistic sharing feature of ISAC channels. The observed spatial congruence between the two channels demonstrates the potential to realize mutual auxiliary functions in ISAC systems.

Although the above efforts have been done to investigate the sharing feature in ISAC systems, this feature has not been incorporated into ISAC channel modeling. Most state-of-the-art ISAC channel models simply perform the generation procedure twice under the consistent framework for communication and sensing channels, which leads to the independence between both channels. This approach contradicts the practical channels with the sharing feature, where some clusters should co-exist or disappear in both channels. Therefore, a realistic ISAC channel model considering the sharing feature is still lacking in the literature, and the corresponding clustering algorithms to extract those shared clusters have not been explored. Moreover, model validation based on practical channel measurements is required to examine the practicality and effectiveness of the channel model for ISAC channel simulations. 

To bridge the above gaps, we have conducted an indoor ISAC channel measurement campaign at 28 GHz and thoroughly investigated the shared relationship between the communication and sensing channels. Our major contributions and novelties are summarized as:

\begin{itemize}
\item{An ISAC channel measurement campaign with horn antenna rotation is performed in typical Line Of Sight (LOS) and Non-LOS (NLOS) indoor scenarios at 28 GHz. The shared scatterers of ISAC channels can be intuitively observed by comparing the Power-Angular-Delay Profiles (PADPs) of communication and sensing measured multipath components (MPCs).}

\item{A shared cluster-based stochastic model is proposed to capture the observed sharing feature of the ISAC channels. Specifically, the shared clusters (contributed by the shared scatterers) and the non-shared ones are defined and superimposed, resulting in the joint generation of communication and sensing channels. Besides, a corresponding clustering algorithm operating on the joint channels is novelly introduced, which aims to extract those shared and non-shared clusters for the model parameterization.}

\item{Stochastic cluster parameters, including the Delay Spread (DS), Angle Spread (AS), and Sharing Degree (SD), are extracted from the measurement data for the model parameterization. In addition, the model empirical implementations are performed, validating the capability to realistically capture the sharing feature.}
\end{itemize}

The remainder of this paper is outlined as follows. In Section \ref{section2}, detailed descriptions of the measurement facilities and environment are presented, and the characteristics of shared scatterers are analyzed. In Section \ref{section3}, a shared cluster-based ISAC channel model and a novel joint clustering algorithm are proposed. Then, parameterized analysis and validation of the proposed channel model are accomplished in Section \ref{section4}. Finally, Section \ref{section5} concludes the work.

\section{Channel Measurements}\label{section2}

\subsection{Measurement Description}

\begin{figure}[h]
\centering
\subfloat[]{\includegraphics[width=3.4in]{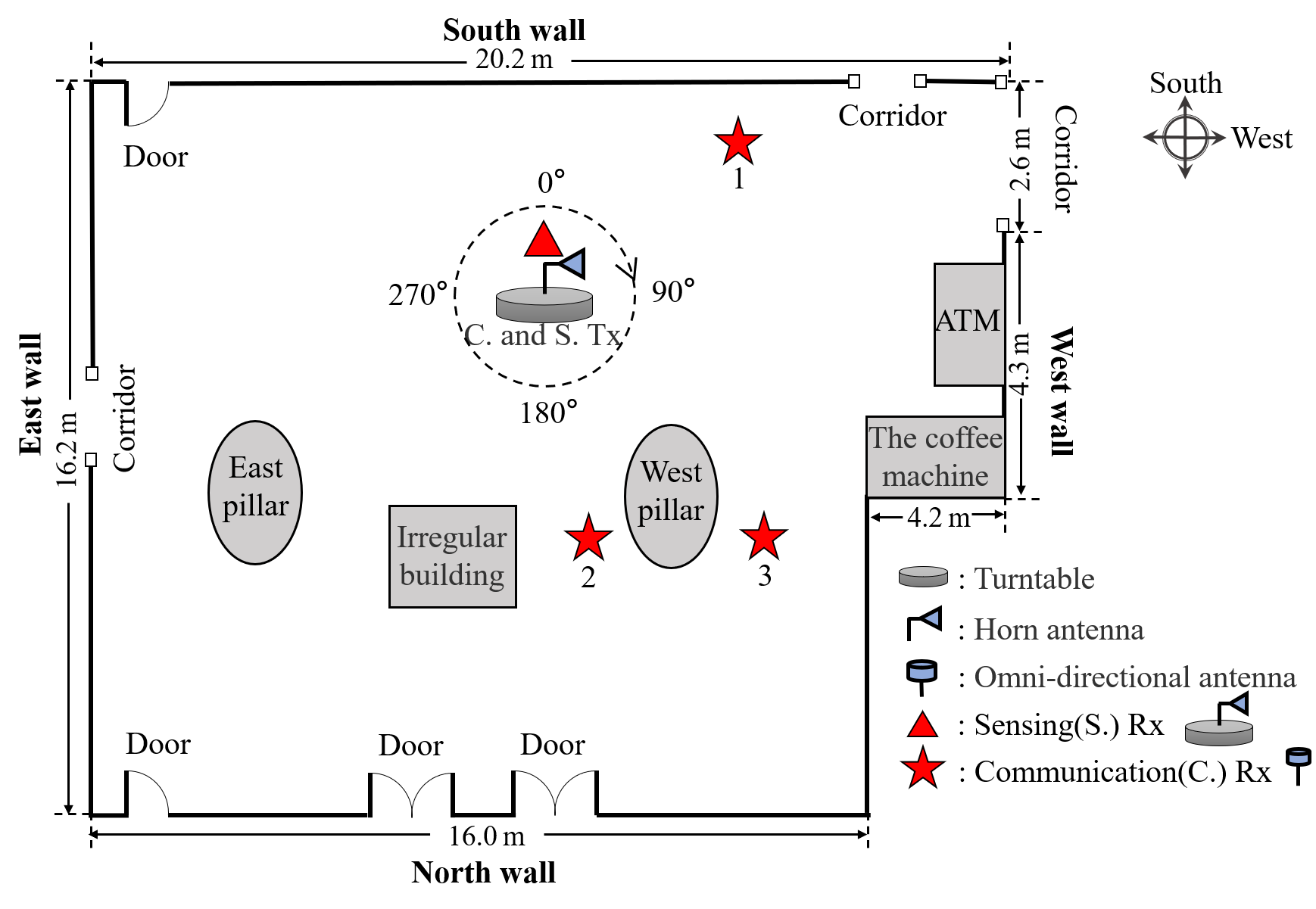}\label{fig_3}}\\
\subfloat[]{\includegraphics[width=2.8in]{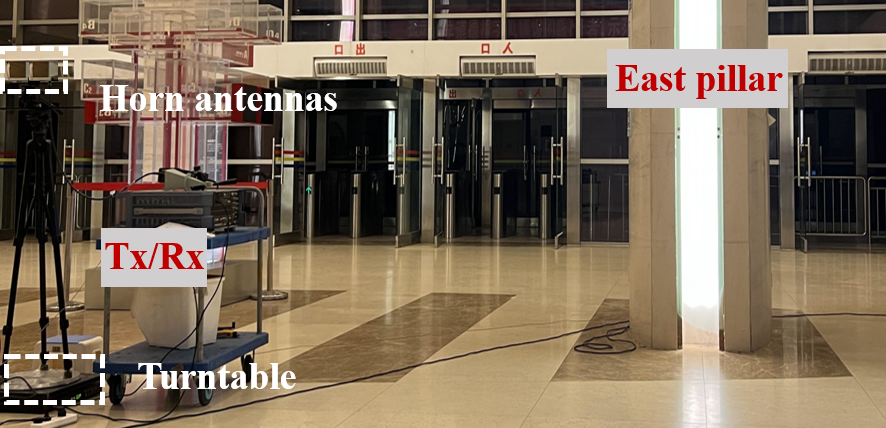}\label{fig_2a}}\\
\subfloat[]{\includegraphics[width=2.8in]{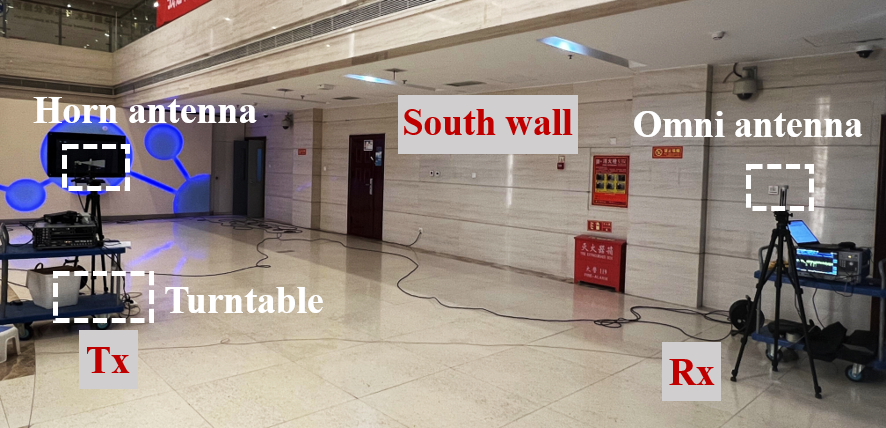}\label{fig_2b}}
\caption{The illustration of the measurement hall layout (a) and measurement photographs taken in (b) the sensing scenario at angle 255$^\circ$ and (c) the communication scenario in position 1 at angle 35$^\circ$.}
\label{fig_2}
\end{figure}

The measurements for communication and sensing channels are conducted in a typical indoor hall with the dimension of 20.2$\times$16.2 $\text{m}^2$ at Beijing University of Posts and Telecommunications. The measurement scenario layout and the realistic measurement surroundings are illustrated in Fig. \ref{fig_2}.

In the sensing measurements (Fig. \ref{fig_2a}), the TX and RX sides are equipped with a horn antenna, respectively. These two antennas are fixed on a bracket with a horizontal interval of 10 cm, pointing in the same direction to detect the round-trip propagation path of sensing signals. The bracket is equipped on a turntable located at the center of the hall, as denoted by the circular dashed area in Fig. \ref{fig_3}. The height of TX/RX antenna is set to 1.47 m. In order to extract the omni-directional channel information, the horn antennas are horizontally rotated clockwise from the south. Therefore, the rotated angles range from 0$^\circ$ to 360$^\circ$ with an increment of 5$^\circ$, resulting in the channel being measured from 72 angles. In the communication measurements, as shown in Fig. \ref{fig_2b}, a rotated horn antenna is still applied as the TX antenna, positioned at the center of the hall. Moreover, the RX side employs an omni-directional antenna, which can receive reflected waves from all directions. This RX antenna is placed in three measured positions respectively, as presented by the red star in Fig. \ref{fig_3}. Positions 1 and 2 are at the LOS condition, while position 3 is at the NLOS condition. 

Note that in this paper, the mono-static sensing equipment is assumed to be integrated into the communication base station for ISAC deployment. This scenario is based on the practical consideration of the base station hardware possessing powerful sensing capabilities. In our work, the communication measurements are conducted with only the TX side rotated. Considering the setup of mono-static sensing, this work is sufficient and time-saving for preliminary investigating and modeling of the sharing feature. The observation and analysis in this paper can be extended for future modeling work based on double rotation measurements.

A wideband channel sounder at mmWave bands is exploited to extract the ISAC channel characteristics. At the TX side, a vector signal generator (R$\&$S SMW 200A) is used to generate a Pseudo Noise (PN) sequence with a code rate of 500 Msym/s and a length of 511. The PN sequence at the baseband is then modulated as a mmWave probing signal at 28 GHz using Binary Phase Shift Keying (BPSK) modulation. The signal has a zero-to-zero bandwidth of 1 GHz. To ensure a better Signal-to-Noise Ratio (SNR) for the signal receiving, a power amplifier with 25 dBm saturation power and 35 dB gain is applied. Then, a high-gain horn antenna is equipped to transmit the amplified signal over the air. At the RX side, a spectrum analyzer (R$\&$S FSW 43) is utilized to demodulate the mmWave signal received by the antenna. 1022 IQ samples are obtained with a sample rate of 1 GHz, and the delay resolution is 1 ns. Finally, the Channel Impulse Responses (CIRs) of the measurement scenarios can be derived through data processing on the laptop.

\begin{table}[h]
\caption{The Channel Sounder Configuration\label{table_1}}
\centering
\begin{tabular}{cc}
\hline
Parameter & Values\\
\hline
Central frequency & 28 GHz\\
Symbol rate & 500 Msym/s\\
Bandwidth & 1 GHz\\
PN sequence & 511\\
Sampling rate & 1 GHz\\
Sensing TX / RX antenna type & Horn / Horn\\
Communication TX / RX antenna type & Horn / Omni\\
Horn antenna azimuth HPBW  & 10$^\circ$\\
Horn antenna gain & 25 dBi\\
Omni-directional antenna gain & 3 dBi\\
Antenna height & 1.47 m\\
\hline
\end{tabular}
\end{table}

In the channel measurements, multiple channel samplings are realized, which are expressed as sampling cycles (snapshots). Each snapshot has a maximum detectable delay of 1022 ns, corresponding to a maximum sounding distance of 306.6 m. Moreover, we synchronize the spectrum analyzer and the signal generator by direct cable connection during the measurements. Details of the channel sounder configuration are listed in Table \ref{table_1}.

\begin{figure*}[h]
\centering
\subfloat[]{\includegraphics[width=3.55in]{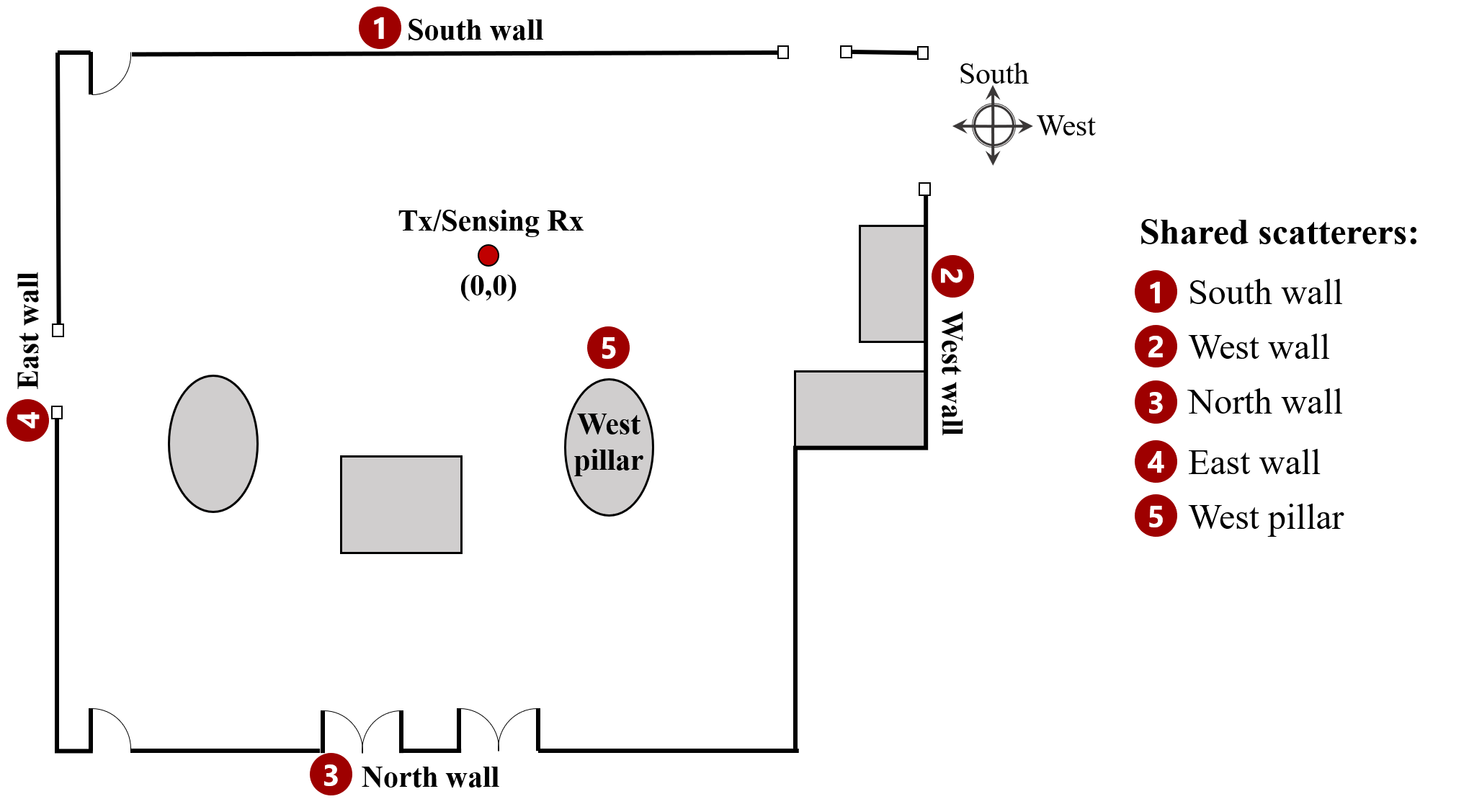}
\label{fig_4legend}}
\subfloat[]{\includegraphics[width=2.33in]{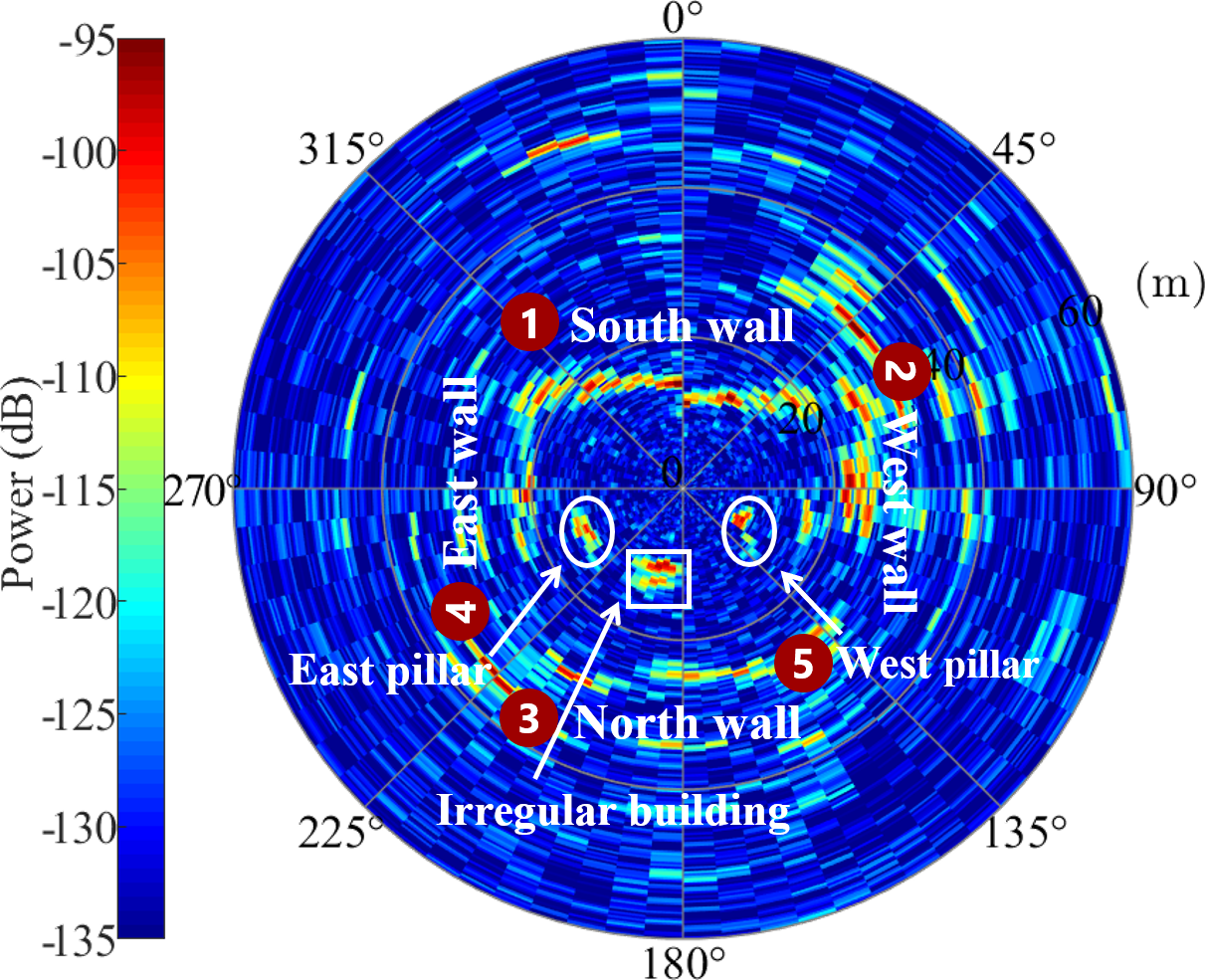}
\label{fig_4a}}\\
\subfloat[]{\includegraphics[width=2.33in]{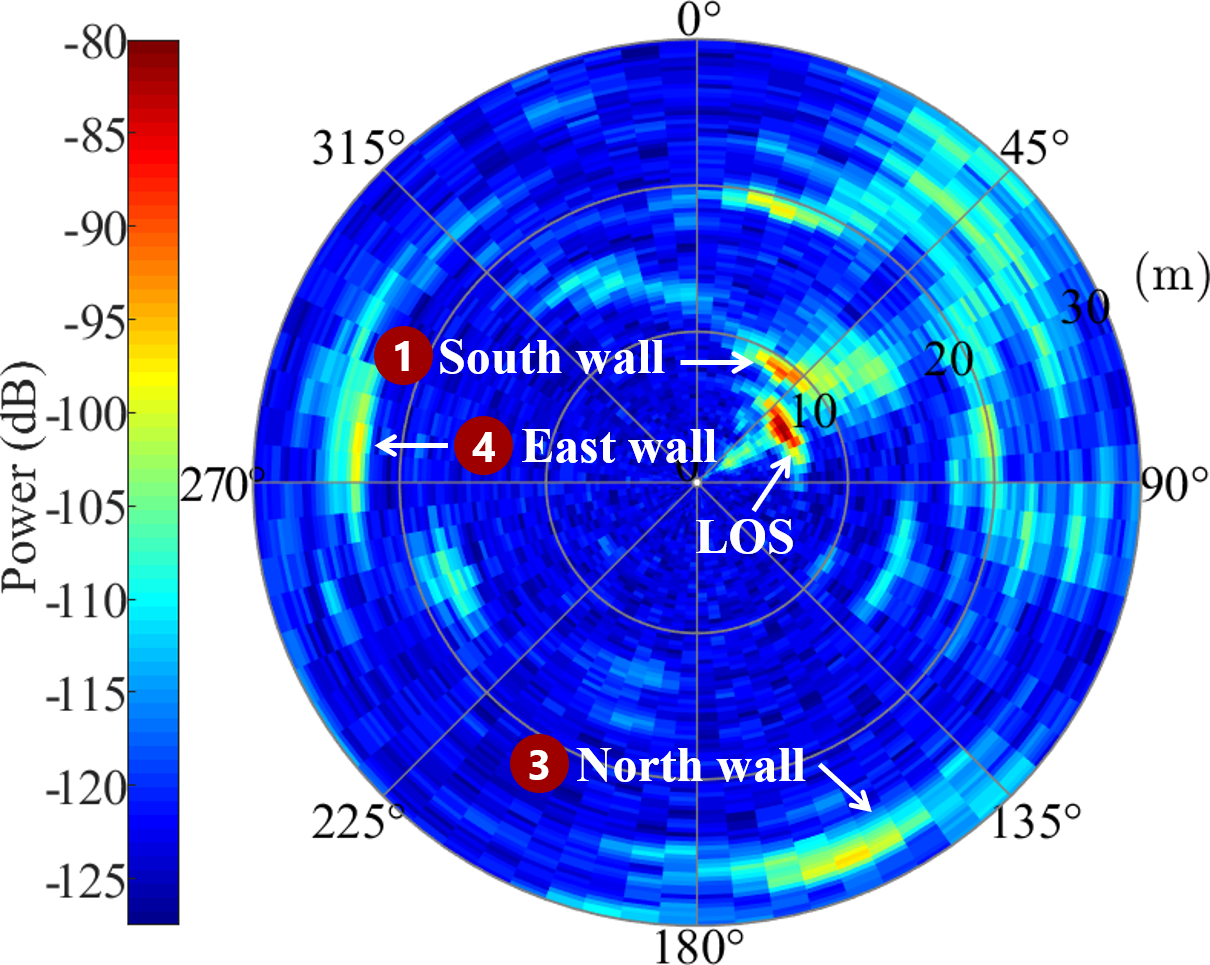}
\label{fig_4b}}
\hfill
\subfloat[]{\includegraphics[width=2.33in]{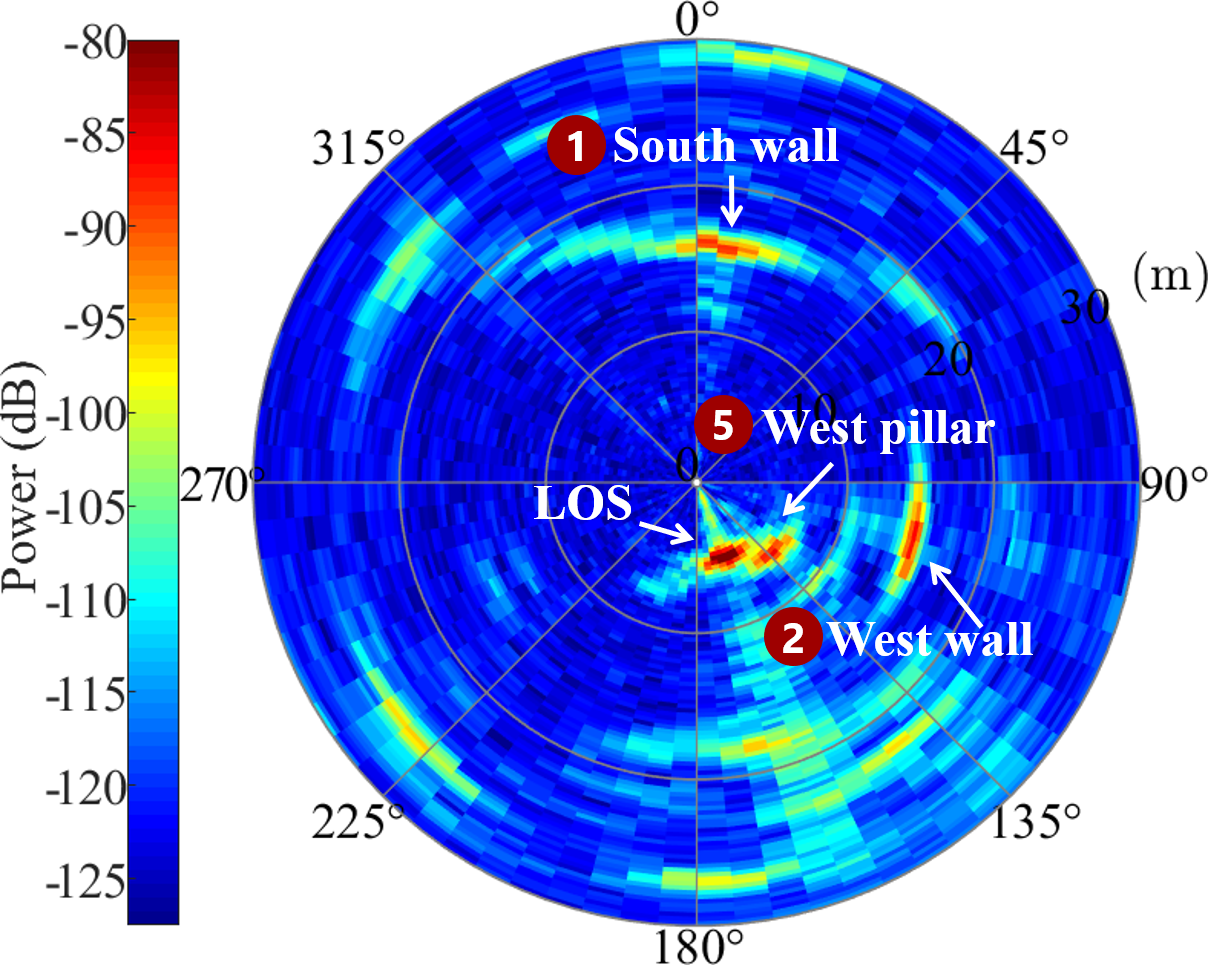}
\label{fig_4c}}
\hfill
\subfloat[]{\includegraphics[width=2.33in]{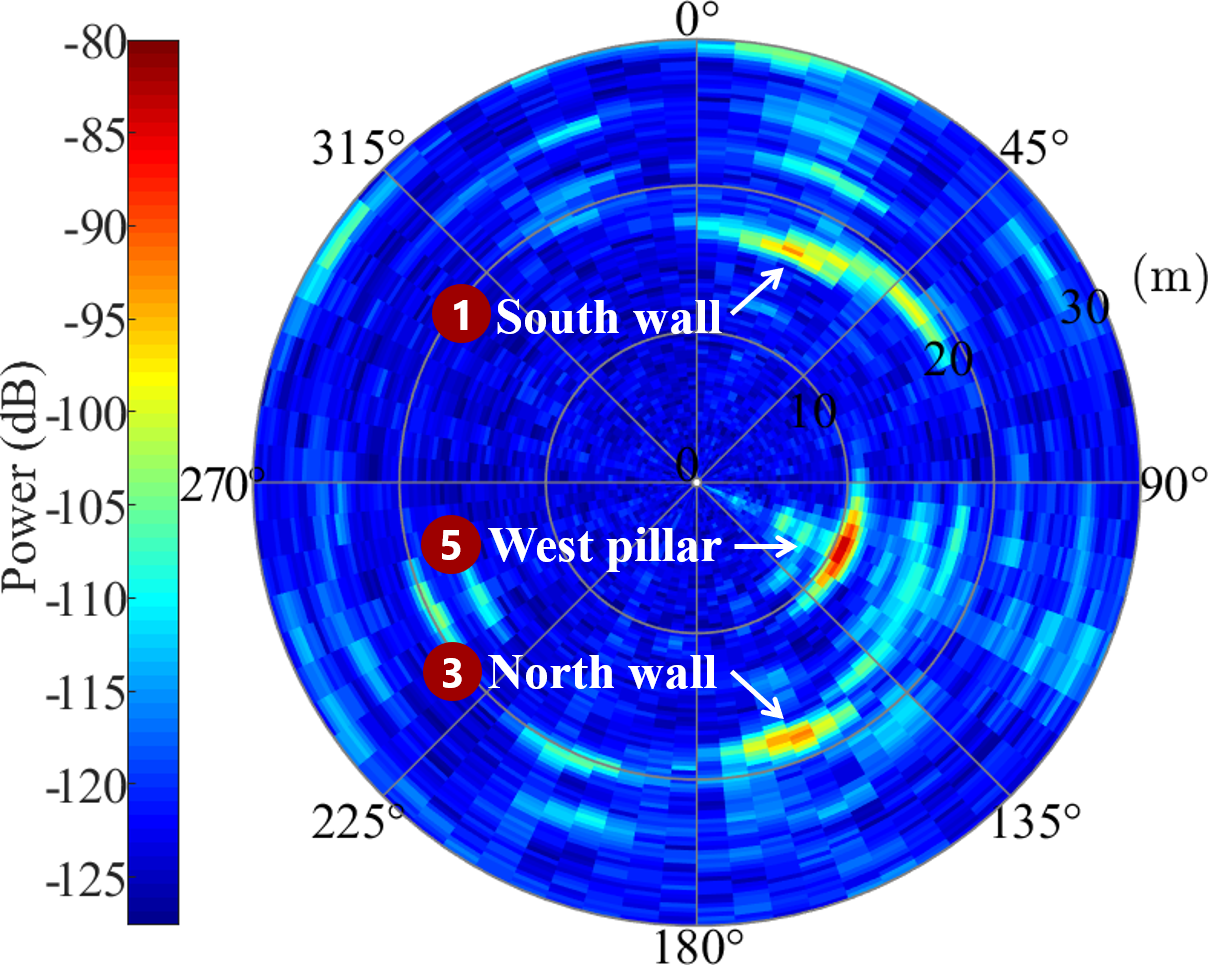}
\label{fig_4d}}
\caption{(a) Schematic of shared scatterers. Measured $\text{PADP}(\theta,\tau)$ in (b) the sensing scenario, and the communication scenarios where the RX antenna is at (c) position 1, (d) position 2, and (e) position 3. Positions 1 and 2 are at the LOS condition. Position 3 is at the NLOS condition. The shared scatterers are marked with red sequence labels.}
\label{fig_4}
\end{figure*}

\subsection{PADP Analysis} 
Before conducting measurements, we perform a back-to-back calibration to remove the system response from the equipment and cables \cite{jiang2019comparative}. CIR at each communication or sensing position for all rotation angles is measured as $h(\theta,\tau)$, where $\tau$ and $\theta$ denote the propagation delay and rotation angle, respectively. The PADPs can be written as

\begin{equation}
\label{eqn_2}
\text{PADP}(\theta,\tau)=|h(\theta,\tau)|^2.
\end{equation}

The PADPs of sensing and three communication measured positions are demonstrated in Fig. \ref{fig_4a}, \ref{fig_4b}-\ref{fig_4d}, respectively. In these polar plots, the center of the circle (coordinate of $(0,0)$) represents the location of the TX and sensing RX antennas, as shown in Fig. \ref{fig_4legend}, which is a simplified schematic of Fig. \ref{fig_3}. The angle of the circle represents the azimuth Angle Of Departure (AOD), with zero-degree pointing in the south direction. The radius indicates the absolute propagation distance from the TX antenna, and the depth of color represents the magnitude of the received power (dB). (Note that the power is not normalized here.) By matching the high-power MPCs with the actual environment, it can be calculated that over 95\% of the MPC power comes from the single-hop reflection, within the 30 dB dynamic range of communication and sensing PADPs.

Based on the characteristics of single-hop reflections, it can be analyzed that the MPCs with similar AOD are contributed by the same propagation scatterer in both communication and sensing PADPs. Moreover, considering the round-trip characteristics of sensing paths, the propagation distance obtained through measurements is twice the actual distance between the scatterers and the TX/RX antenna. As a result, the absolute delay and AOD of sensing MPCs can be used to determine the location of environmental objects, enabling localization. As demonstrated in Fig. \ref{fig_4a}, the scenario distribution, including the walls around the hall, the east and west pillars, and the irregularly shaped building, is presented clearly, which can be corresponded to Fig. \ref{fig_3}. However, at the location of corridors and doors (e.g., in the area around angle 45$^\circ$ in Fig. \ref{fig_4a}), MPCs are randomly distributed due to multiple reflections, diffractions, etc., caused by irregular scatterers.

In communication measurements, the delay parameter is associated with the overall signal propagation from TX to RX. For single-hop paths at the NLOS condition, the distance between the TX antenna and the scatterer can be geometrically calculated based on the delay and angle. However, for multi-hop paths, the distance cannot be determined. Instead, the angle between the TX antenna to the first-hop scatterer is always intuitively reflected by the AOD of MPCs in the PADP. Based on the above analysis, it can be observed from angular domain that the high-power MPCs in communication PADP 1 (as illustrated in Fig. \ref{fig_4b}) are associated with partial south, north, and east walls, with the LOS path exhibiting the highest power. Moreover, the primary channel scatterers, corresponding to high-power MPCs, are also marked in the PADPs of communication position 2 (Fig. \ref{fig_4c}) and position 3 (Fig. \ref{fig_4d}).

Upon analyzing the PADPs of communication and sensing channels, it becomes evident that some environmental objects (marked with red sequence labels in Fig. \ref{fig_4}) are not only the sensing targets but also assist in facilitating communication transmission. (Note that some objects, such as the north wall, may have large physical sizes. However, only certain parts of these objects can effectively scatter the multipath of communication and sensing signals, while other parts of these objects may not contribute to the received signals at the RX.) For instance, the partial south wall (label 1), north wall (label 3), and east wall (label 4) when the communication RX is set in position 1 (compared Fig. \ref{fig_4a} with Fig. \ref{fig_4b}). In this paper, we define these scatterers as shared scatterers of ISAC channel. Compared Fig. \ref{fig_4a} with Fig. \ref{fig_4c}, the partial south wall (label 1), west wall (label 2), and west pillar (label 5) become shared scatterers in channels. Compared Fig. \ref{fig_4a} with Fig. \ref{fig_4d}, shared scatterers are the partial south wall (label 1), north wall (label 3), and west pillar (label 5). While communication and sensing MPCs naturally exhibit different propagation distances and angular deviations due to variations in beam widths and TX-RX relative positions, this does not negate the fact that the objects act as shared propagation scatterers for both communication and sensing channels.

Here, the sensing channel incorporates all environmental scatterers (i.e., waves propagating to the scatterers will always return a portion of the energy that can be received), due to the high detection dynamic range provided by the horn antenna gain and the powerful diffuse reflection mechanisms facilitated by the rough surfaces. Therefore, the shared scatterers of three ISAC channels can be intuitively observed in communication PADPs (Fig. \ref{fig_4b}-\ref{fig_4d}), respectively. These observed shared scatterers (corresponding to the MPCs) reflect the realistic propagation paths of the electromagnetic waves, revealing the sharing feature of ISAC channels. This is a practical environmental characteristic that cannot be ignored when modeling the channel. 

\begin{figure*}[!h]
\centering
\includegraphics[width=5.5in]{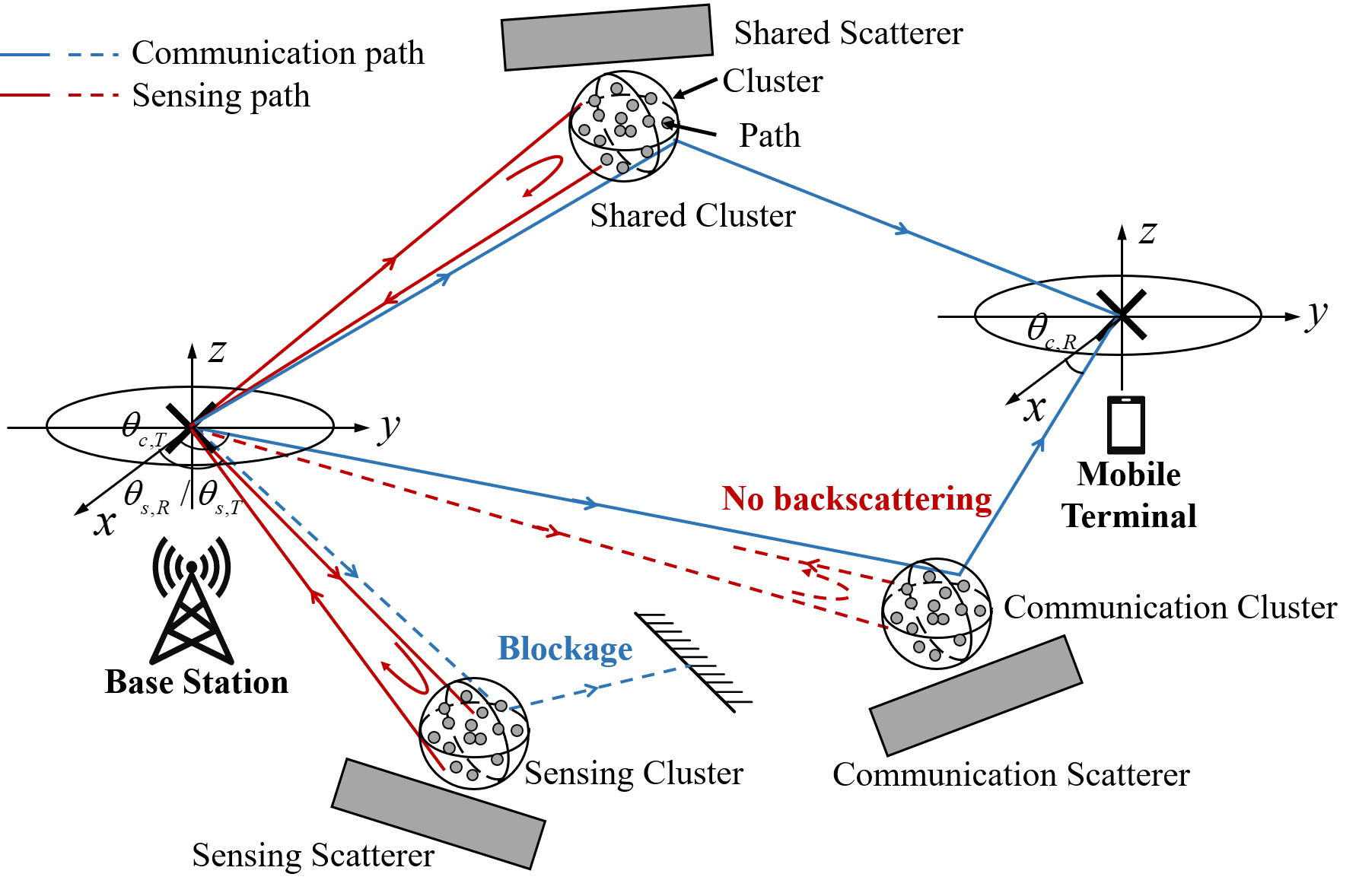}
\caption{The illustration of ISAC channel model. The blue lines denote communication paths, and the red lines denote sensing paths. The dashed lines indicate that the paths are interrupted.}
\label{fig_1}
\end{figure*}

\begin{figure*}
\begin{subequations}
\label{eqn_1}
\begin{align}
\label{eqn_1a}
h_{c}(\theta_{c,R},\theta_{c,T}, \tau_{c})=&\underbrace{\sum\limits_{n_0=1}^{N_0}\sum\limits_{m_0=1}^{M_0} a_{c,n_0,m_0}\delta(\theta_{c,R}-\theta_{c,n_0,m_0,R})\delta(\theta_{c,T}-\theta_{c,n_0,m_0,T})\delta(\tau_{c}-\tau_{c,n_0,m_0})}_\text{Shared\ Communication\ Sub-Clusters}\notag\\&+\sum\limits_{n_1=1}^{N_1}\sum\limits_{m_1=1}^{M_1} a_{c,n_1,m_1}\delta(\theta_{c,R}-\theta_{c,n_1,m_1,R})\delta(\theta_{c,T}-\theta_{c,n_1,m_1,T})\delta(\tau_{c}-\tau_{c,n_1,m_1}),\\
\label{eqn_1b}
h_{s}(\theta_{s,R},\theta_{s,T}, \tau_{s})=&\underbrace{\sum\limits_{n_0=1}^{N_0}\sum\limits_{m_0=1}^{M_0} a_{s,n_0,m_0}\sigma_{n_0,m_0}\delta(\theta_{s,R}-\theta_{s,n_0,m_0,R})\delta(\theta_{s,T}-\theta_{s,n_0,m_0,T})\delta(\tau_{s}-\tau_{s,n_0,m_0})}_\text{Shared\ Sensing\ Sub-Clusters}\notag\\&+\sum\limits_{n_2=1}^{N_2}\sum\limits_{m_2=1}^{M_2} a_{s,n_2,m_2}\sigma_{n_2,m_2}\delta(\theta_{s,R}-\theta_{s,n_2,m_2,R})\delta(\theta_{s,T}-\theta_{s,n_2,m_2,T})\delta(\tau_{s}-\tau_{s,n_2,m_2}).
\end{align}
\end{subequations}
\end{figure*}

\section{ISAC Channel Modeling}\label{section3}

In this section, we introduce the proposed shared cluster-based channel model for ISAC systems, as shown in Fig. \ref{fig_1}, which encompasses the parameters of shared and non-shared clusters. The SD metric is proposed after the model descriptions. In addition, to extract the defined shared and non-shared clusters, a clustering algorithm operating on the ISAC channels is introduced.

\subsection{Channel Modeling}

This paper considers a wideband ISAC channel with one TX antenna and one RX antenna, under frequency-selective fading. As shown in Fig. \ref{fig_1}, the gray squares represent randomly distributed scatterers. Each circle with several dots represents one scattering region (cluster) causing one group of paths with similar properties \cite{zhang20173}. In the conventional configuration, the communication involves the transmission of signals from the base station, represented by the blue lines in Fig. \ref{fig_1}, which are received by the mobile terminal after experiencing the channel scatterers. Meanwhile, the base station senses the surrounding environment through the echo signals, which are reflected back to itself for reception, as shown by the red lines in Fig. \ref{fig_1}. It should be noted that the model in this paper simplifies the consideration of some issues, such as the vertical angles and Multiple-Input and Multiple-Output (MIMO) antenna elements, in comparison to ITU-R M.2412 \cite{itu2412}. This simplification allows for a focused exploration of the sharing feature in ISAC channels. Nevertheless, the model remains applicable even when incorporating or extending these factors.

We define the clusters that experience the shared scatterers in both the communication and sensing channels as shared clusters of the ISAC channel. In our proposed ISAC channel model, the communication and sensing channels are generated by the superposition of shared and non-shared clusters, respectively. The channel model of all resolvable paths at time $t$ is expressed as formula (\ref{eqn_1}). In this model, $h_{c}(\theta_{c,R},\theta_{c,T}, \tau_{c})$ and $h_{s}(\theta_{s,R},\theta_{s,T}, \tau_{s})$ denote the CIRs of communication and sensing, which are the general expression expanded $h(\theta,\tau)$ in Section \ref{section2}. 

\begin{itemize}
\item{The subscripts, $c$ and $s$, denote the communication and sensing propagations, respectively.}

\item{The subscript $T$ stands for the transmitting propagation, and $R$ stands for the receiving propagation.} 

\item{$\{1,\dots,n_0,\dots,N_0\}$, $\{1,\dots,n_1,\dots,N_1\}$, and $\{1,\dots,n_2,\dots,N_2\}$ respectively denote shared, communication, and sensing clusters, where $n_0$, $n_1$, $n_2$, and $N_0$, $N_1$, $N_2$ represent the indexes and total numbers. The total number of communication and sensing clusters can be given as $N_c=N_0+N_1$ and $N_s=N_0+N_2$, and the joint cluster number is $N=N_0+N_1+N_2$.}

\item{$m_0$, $m_1$, $m_2$, and $M_0$, $M_1$, $M_2$ denote the indexes and total numbers of the MPCs within the corresponding clusters.}

\item{$\delta(\cdot)$ denotes the Dirac Delta function.}

\item{$a_{c,n_0,m_0}$, $a_{c,n_1,m_1}$, $a_{s,n_0,m_0}$, and $a_{s,n_2,m_2}$ represent the amplitude of cluster $n_0^{th}$ MPC $m_0^{th}$ and cluster $n_1^{th}$ MPC $m_1^{th}$ in communication CIR, and cluster $n_0^{th}$ MPC $m_0^{th}$ and cluster $n_2^{th}$ MPC $m_2^{th}$ in sensing CIR, respectively.}

\item{$\theta_{c,n_0,m_0,R}$, $\theta_{c,n_1,m_1,R}$, $\theta_{s,n_0,m_0,R}$, and $\theta_{s,n_2,m_2,R}$ are the azimuth Angle Of Arrival (AOA) of cluster $n_0^{th}$ MPC $m_0^{th}$ and cluster $n_1^{th}$ MPC $m_1^{th}$ in communication CIR, and cluster $n_0^{th}$ MPC $m_0^{th}$ and cluster $n_2^{th}$ MPC $m_2^{th}$ in sensing CIR, respectively.}

\item{$\theta_{c,n_0,m_0,T}$, $\theta_{c,n_1,m_1,T}$, $\theta_{s,n_0,m_0,T}$, and $\theta_{s,n_2,m_2,T}$ are the azimuth AOD of cluster $n_0^{th}$ MPC $m_0^{th}$ and cluster $n_1^{th}$ MPC $m_1^{th}$ in communication CIR, and cluster $n_0^{th}$ MPC $m_0^{th}$ and cluster $n_2^{th}$ MPC $m_2^{th}$ in sensing CIR, respectively.}

\item{$\tau_{c,n_0,m_0}$, $\tau_{c,n_1,m_1}$, $\tau_{s,n_0,m_0}$, and $\tau_{s,n_2,m_2}$ represent the delay of cluster $n_0^{th}$ MPC $m_0^{th}$ and cluster $n_1^{th}$ MPC $m_1^{th}$ in communication CIR, and cluster $n_0^{th}$ MPC $m_0^{th}$ and cluster $n_2^{th}$ MPC $m_2^{th}$ in sensing CIR, respectively.}

\item{$\sigma_{n_0,m_0}$ and $\sigma_{n_2,m_2}$ denote the Radar
Cross Section (RCS) coefficient of sensing cluster $n_0^{th}$ MPC $m_0^{th}$ and cluster $n_2^{th}$ MPC $m_2^{th}$, respectively.}

\end{itemize}

The purpose of communication is to transfer information without concern to the environmental scatterers. On the other hand, sensing aims at the detection and identification of environmental objects. Therefore, when modeling the sensing CIRs, it is essential to further describe the properties of the objects. One important parameter to consider is the RCS, which is influenced by factors such as object area and material. The inclusion of RCS has been studied in the research on sensing channels \cite{noh2022communication}. In model (\ref{eqn_1}), the RCS coefficient, denoted as $\sigma$, is adopted to characterize the fading of the cluster power due to sensing scatterers. Moreover, the shared clusters in communication and sensing channels are represented as shared sub-clusters, which are marked with braces in model (\ref{eqn_1}). The shared sub-clusters of communication and sensing channels differ in certain parameters. While the angles of the sub-cluster centroids are approximately consistent, the delays should be considered separately or mathematically calculated according to the TX-RX positions.

The physical propagation paths experienced by reflection, scattering, diffraction, etc., are subject to complex environmental conditions, and not all objects contribute to both communication and sensing channels. As illustrated in Fig. \ref{fig_1}, some sensing scatterers will only contribute sensing clusters as denoted in model (\ref{eqn_1}), when they are in the vicinity of the TX (i.e., the base station) and the path to the communication RX (i.e., the mobile terminal) is blocked. On the other hand, only the specific directed paths to the mobile terminal are generated and with difficulty establishing a backscattering path to the base station in the following potential circumstances: The scatterer surface exhibits significant specular properties. The angle between the incident electromagnetic wave and the scatterer surface is too small, such as in narrow corridor scenes. The scatterer is close to the communication RX and affects the communication channels but is challenging to be detected due to the low echo power of the sensing process. In the above cases, these scatterers will only contribute to communication clusters as denoted in model (\ref{eqn_1}).

In this section, only the dominant single-hop paths are primarily considered, and retaining multi-hop analysis will result in a more complex sharing relationship for ISAC channels. Moreover, it should be noted that this paper focuses on the omni-directional channels of communication and sensing. When communication and sensing beams are directed to different regions through beamforming \cite{kumari2021adaptive, gonzalez2016radar}, shared and non-shared scatterers and clusters appear as a result of the beam coverage.  

To measure the sharing feature of the ISAC channels, we define the Sharing Degree (SD) metric, which measures the power ratio of the shared clusters in communication and sensing channels, respectively. Based on the proposed model, the sensing SD is expressed as

\begin{align}
\text{SD}_{s}&=\dfrac{P_{s}^{\text{shared}}}{P_{s}^{\text{total}}}\notag\\&=\dfrac{ \left|\sum\limits_{n_0}\sum\limits_{m_0} a_{s,n_0,m_0}\sigma_{n_0,m_0} \right|^2}{\left|\sum\limits_{n_0}\sum\limits_{m_0} a_{s,n_0,m_0}\sigma_{n_0,m_0}+\sum\limits_{n_2}\sum\limits_{m_2} a_{s,n_2,m_2}\sigma_{n_2,m_2}\right|^2},
\end{align}
where $P_s^{\text{shared}}$ and $P_s^{\text{total}}$ are the power of shared sensing sub-clusters and total clusters ($N_s$) in model (\ref{eqn_1}), respectively. Similarly, the communication SD is expressed as $\text{SD}_{c}=P_{c}^{\text{shared}}/P_{c}^{\text{total}}$. In these formulas, $P^{\text{shared}}/P^{\text{total}}$ stands for the proportion of shared parts in the respective received power. 

\subsection{Joint Clustering for Communication and Sensing Channels}

The clustering technique is crucial as it fulfills the need to group MPCs with similar channel characteristics for subsequent modeling. A review on existing channel clustering algorithms and their performance evaluation is performed in \cite{huang2022artificial}. However, the shared clusters in the proposed model are contributed by both the communication and sensing channels. To extract those clusters (shared and non-shared ones), a joint clustering with both channels integrated is necessary. In this part, a novel KPowerMeans-based Joint Clustering Algorithm (KPM-JCA) is introduced in the following to address this issue. 

Firstly, the PADPs of the measured sensing and communication channels are denoised with the dynamic range of 30 dB from the peak value, to ensure the consistency of MPC selection and remove the multi-reflection paths with low power \cite{ma2017data}. All the positive data after denoising are selected as effective MPCs, expressed by

\begin{equation}\label{eqn_3}
path_{c/s}=\{\theta_m,\tau_m,p_m\},m=\{1,2,\cdots,M\},
\end{equation}
where $m$ and $M$ are the index and the total number of valid MPCs in the measurements, respectively. $\theta$ and $\tau$ have the same meanings as in formula (\ref{eqn_2}). Note that the azimuth AOD of MPCs would be estimated directly according to the rotation angle. $p_m=|a_m|^2$ denotes the power of the $m^{th}$ MPC.

\begin{figure}[h]
\centering
\subfloat[]{\includegraphics[width=3.3in]{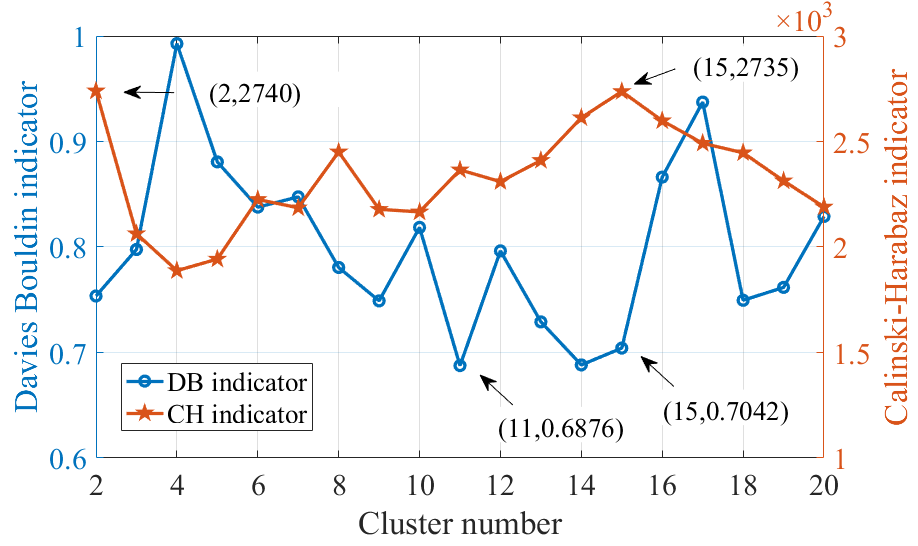}\label{fig_9a}}\\
\subfloat[]{\includegraphics[width=3.3in]{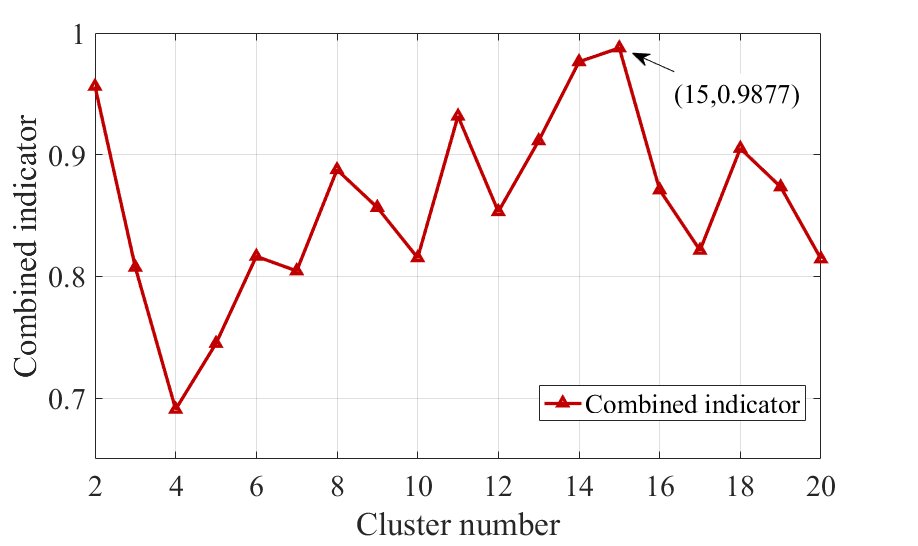}\label{fig_9b}}
\caption{Clustering evaluation by (a) DB and CH indicators, and (b) combined indicator, under the different cluster number.}
\label{fig_9}
\end{figure}

Subsequently, in contrast to traditional clustering algorithms employed for a single channel, we combine the MPC parameter sets from the sensing channel ($\{path_{s}\}$) and the communication channel ($\{path_{c}\}$), while also incorporating additional data labeling for the respective sources. This approach enables the joint effective parameter set of ISAC channel being extracted by $\{path_{\text{ISAC}}\}=\{path_{s}\}\cup\{path_{c}\}$. Note that although the functions and devices in ISAC systems are integrated, communication and sensing channels are essentially two links with potentially different propagation power losses. Therefore, this paper introduces a power compensation coefficient, denoted as $\gamma$, for communication MPCs in the subsequent KPowerMeans algorithm. This coefficient aims to balance and normalize the power impact, enabling the effective extraction of the shared clusters that serve both two channels. The calculation is represented as $p_m=p_m\cdot\gamma$, $p_m\in\{path_{c}\}$.

Considering the complexity and practicality of the algorithm, we utilize the Multipath Component Distance (MCD) of the delay and angle dimension to weigh MPC similarity and then apply the classic KPowerMeans algorithm for clustering \cite{czink2006framework,Li2020Clustering}. To improve the certainty and accuracy of clustering results, a step in \cite{huang2009cluster} is adopted to get initial centroid positions with distance checking. Moreover, for automatic clustering, a range for the anticipated cluster number, $K=[K_\text{min}, K_\text{max}]$, should be provided in advance. 

For clustering performance evaluation, the Davies Bouldin (DB) and Calinski-Harabaz (CH) are two well-known indicators \cite{Li2020Clustering,czink2006framework}. The lower value of the DB indicator stands for closer paths within clusters, while the higher value of the CH indicator stands for more separate clusters. According to the backscattered echo propagation, the scatterers (e.g., the walls) in the indoor environment are continuously detected, resulting in the continuous delay and angle of sensing MPCs, as well as insignificant step changes in MPC power. Therefore, the measured MPCs do not have distinct centralized clustering characteristics, as corroborated in Fig. \ref{fig_4a}. This brings challenges to the evaluation of clustering performance and the selection of optimal cluster numbers in ISAC channels. 

Taking the measured data of ISAC channel 1 (where the communication RX antenna is set at position 1) as an example, we set the range of $K$ as $[2:20]$. Fig. \ref{fig_9a} shows the DB and CH values for each cluster number, which are the average results of multiple snapshots. It can be observed that the DB and CH curves are non-convex functions with large volatility. When $K=11$, the optimal value of DB is 0.6876, and when $K=2$, the optimal value of CH is 2740. The optimal numbers of clusters revealed by the two classic indicators are not consistent, and the values of the other indicator are not ideal at the single optimal points.

In the literature, some work has been done to combine these two classic metrics. For instance, a cluster validation method with a DB threshold is proposed in \cite{czink2006framework}, where the value of $K$ that maximizes CH values within the DB threshold is selected as the optimal cluster number. However, considering the fluctuating data of measured ISAC channels, as expounded in Fig. \ref{fig_9a}, even the same dataset with different thresholds will lead to various cluster number selection results. Consequently, to obtain the desired clustering results, the threshold usually needs to be subjectively adjusted based on the specific values, which becomes inefficient when the data amount is large. Therefore, this paper introduces a trade-off method for the DB and CH indicators to automatically evaluate and select the optimal number of clusters. The calculation can be expressed as follows

\begin{equation}\label{eqn_4}
K^*=\arg\max_K\left\{\frac{1}{2}\cdot\left[\frac{\text{DB}_\text{min}}{\text{DB}(K)}+\frac{\text{CH}(K)}{\text{CH}_\text{max}}\right]\right\},
\end{equation}
where $\text{DB}(K)$ and $\text{CH}(K)$ denote the DB and CH values, respectively, when the cluster number is $K$. $\text{DB}_{min}$ and $\text{CH}_{max}$ are the corresponding optimal values, traversing $[K_{min},K_{max}]$. The value range of the combined indicator in $\{\cdot\}$ is $[0,1]$, and the optimal number of clusters is $K^*$, which maximizes the combined indicator.
 
\begin{figure*}[t]
\centering
\subfloat[]{\includegraphics[height=1.94in]{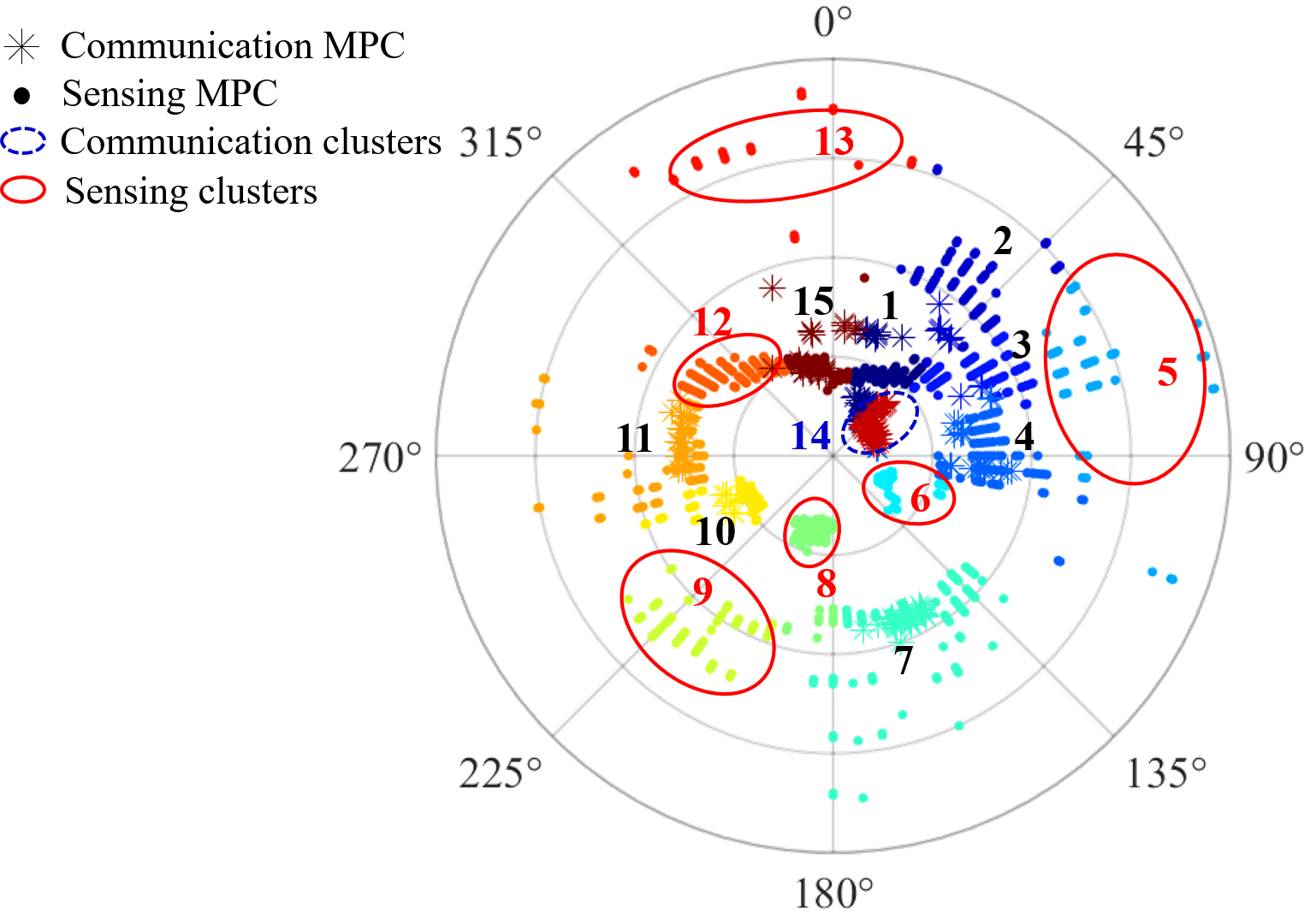}
\label{fig_5a}}
\hfil
\subfloat[]{\includegraphics[height=1.94in]{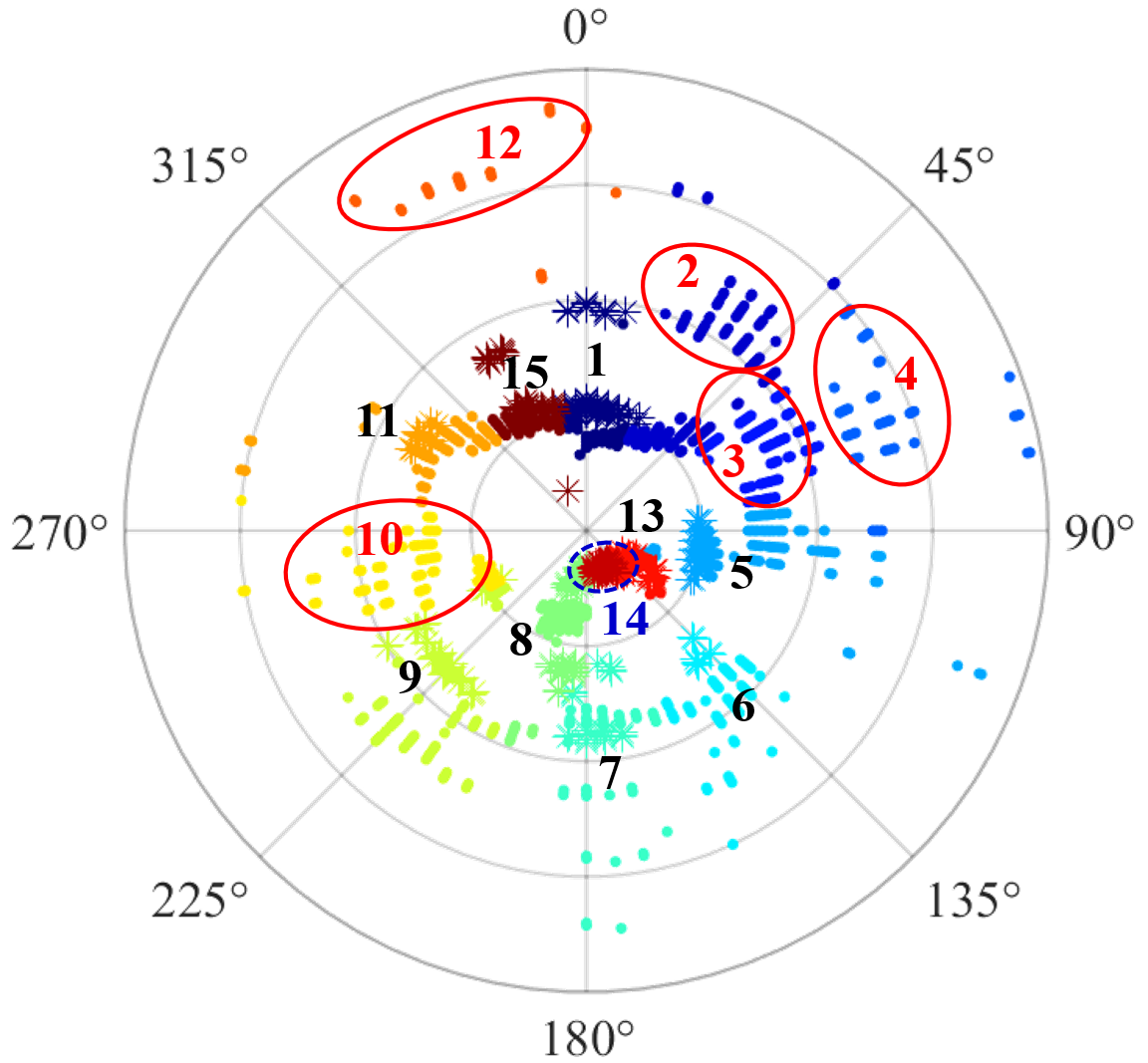}
\label{fig_5b}}
\hfil
\subfloat[]{\includegraphics[height=1.94in]{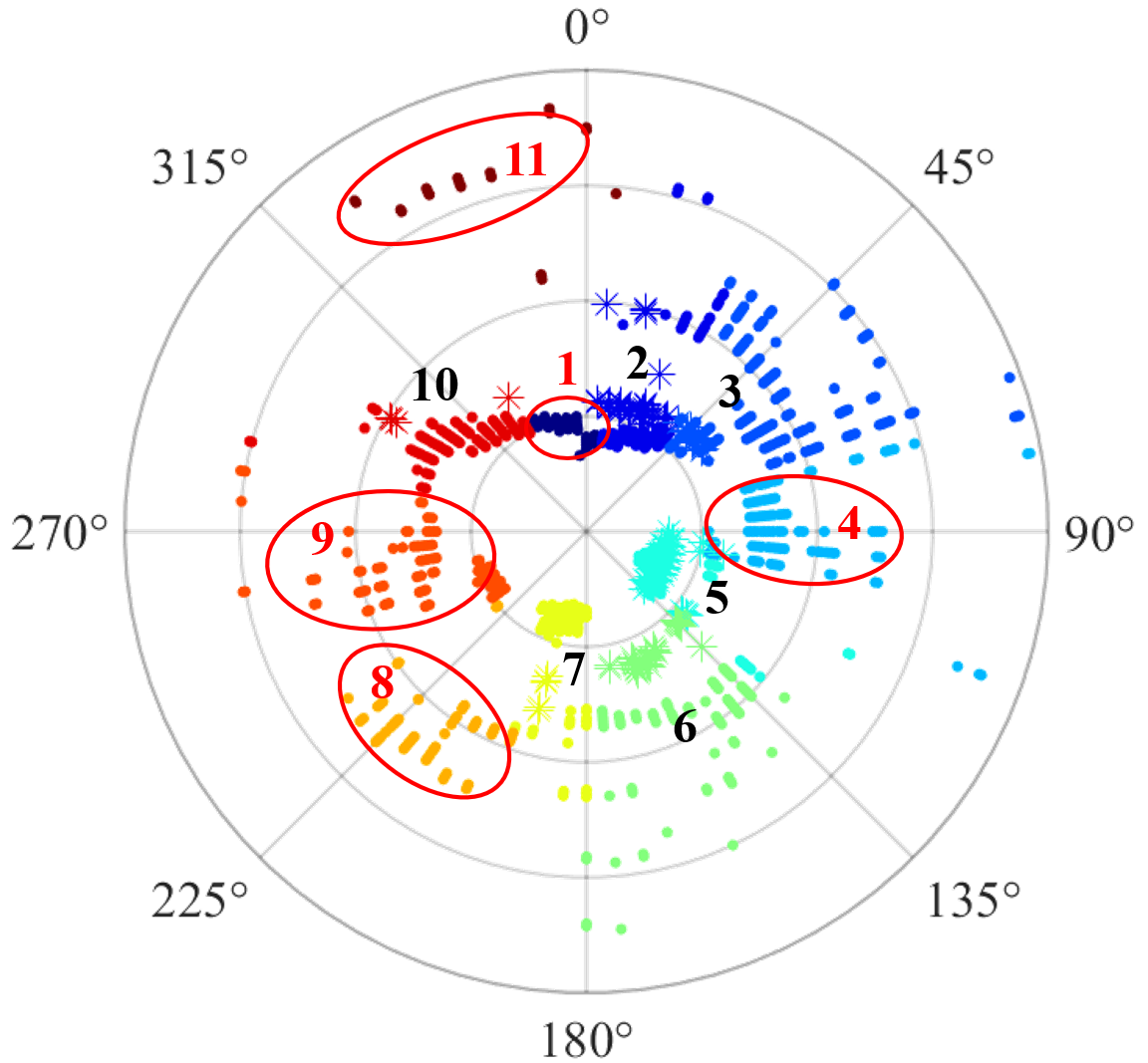}
\label{fig_5c}}
\caption{Schematic of measured clustering results in ISAC channels, where the communication RX antenna is at (a) position 1, (b) position 2, and (c) position 3. Positions 1 and 2 are at the LOS condition. Position 3 is at the NLOS condition. Clusters are distinguished by different serial numbers and colors, with black numbers indicating shared clusters.}
\label{fig_5}
\end{figure*}

The quantified results of ISAC channel 1, obtained by applying the formula (\ref{eqn_4}), are shown in Fig. \ref{fig_9b}. When $K=15$, the combined indicator reaches its optimal value, with sub-optimal values of 0.7042 for the DB indicator and 2735 for the CH indicator (as shown in Fig. \ref{fig_9a}). In this case, there is indeed a more reasonable visual clustering result, which has been presented in subsequent Fig. \ref{fig_5a}. Note that while this trade-off selection method for determining the cluster number is proposed specifically for ISAC channel MPCs with insignificant clustering characteristics, it is also applicable to regular data. 

In summary, the shared and non-shared clusters are automatically extracted for ISAC channel modeling after performing this joint clustering on the given measurements gathered in a given unknown environment. Specifically, the cluster that contains both communication and sensing MPCs in clustering results corresponds to the shared cluster, while the cluster only contains communication or sensing MPCs identifies the communication or sensing cluster.

\section{Model Parameterized Analysis and Validation}\label{section4}

\begin{table*}[t]
\caption{Joint Clustering Results of Measured MPCs\label{table_2}}
\centering
\begin{tabular}{cccccc}
\hline
Channels & Cluster number & Sensing cluster number (label) & Communication cluster number (label) & Shared cluster number (label) & $\text{SD}_s$\\
\hline
1 (LOS) & 15 & 14 (1-13,15) & 9 (1,2,3,4,7,10,11,14,15) & 8 (1,2,3,4,7,10,11,15) & 0.65\\
2 (LOS) & 15 & 14 (1-13,15) & 10 (1,5,6,7,8,9,11,13,14,15) & 9 (1,5,6,7,8,9,11,13,15) & 0.64\\
3 (NLOS) & 11 & 11 (1-11) & 6 (2,3,5,6,7,10) & 6 (2,3,5,6,7,10) & 0.54\\
\hline
\end{tabular}
\end{table*}

Here we present the procedure for generating a realistic ISAC channel based on the proposed modeling framework. The measured data of communication and sensing channels are employed as input for the modeling procedure. The output of this procedure is CIRs for communication ($h_c$) and sensing ($h_s$). Firstly, in Module 1, the MPC parameter sets (denoted as $\{path_{c/s}\}$ in (4)) are extracted. Then, in Module 2, all measured MPCs are clustering, and the cluster statistics such as DS, and AS are obtained for both communication and sensing channels using the proposed KPM-JCA algorithm of Section \ref{section3}-B. Based on these statistics and ITU-R M.2412 standards \cite{itu2412}, shared and non-shared cluster parameters are realized in Module 3, and steps \emph{1)}-\emph{3)} of Section \ref{section4}-C present the implementation. Finally, the shared cluster-based ISAC channel at the small-scale level can be generated in Module 4 according to step \emph{4)} of Section \ref{section4}-C and formula (2). Algorithm \ref{alg:alg1} demonstrates the detailed implementation of the proposed framework.

\begin{algorithm}
\caption{Implementation of the proposed modeling framework for ISAC channel}\label{alg:alg1}
\renewcommand{\algorithmicrequire}{\textbf{Input:}}
\renewcommand{\algorithmicensure}{\textbf{Output:}}
\begin{algorithmic}[1]
\REQUIRE Measured data of communication and sensing channel.
\ENSURE CIRs for communication ($h_c$) and sensing ($h_s$).\\
{\%\%\textbf{Module 1 (1-2):} MPC parameter extraction, Section \ref{section3}-B}
\STATE{Denoise the measured data.}
\STATE{Extract measured MPC parameter set of communication channel ($\{path_{c}\}$) and sensing channel ($\{path_{s}\}$) via (\ref{eqn_3}).\\}
{\%\%\textbf{Module 2 (3-7):} MPC clustering, Section \ref{section3}-B}
\STATE{Generate the joint MPC parameter set via $\{path_{\text{ISAC}}\}=\{path_{s}\}\cup\{path_{c}\}$.}
\STATE{Normalize communication and sensing MPC power with a compensation coefficient $\gamma$ for communication MPCs via $p_m=p_m\cdot\gamma$, $p_m\in\{path_{c}\}$.}
\STATE{Apply KPowerMeans algorithm [28],[29] on $\{path_{\text{ISAC}}\}$ for clustering the joint MPCs.}
\STATE{Evaluate and select the optimal clustering result via (\ref{eqn_4}).}
\STATE{Obtain the stochastic DS, AS, etc.\\}
{\%\%\textbf{Module 3 (8-10):} Cluster parameter generation, Section \ref{section4}-C}
\STATE{Generate parameters of sensing clusters via step \emph{1)}.}
\STATE{Generate shared parameters of communication clusters via step \emph{2)}.}
\STATE{Generate stochastic parameters of communication clusters via step \emph{3)}.\\}
{\%\%\textbf{Module 4 (11):} CIR modeling, Section \ref{section4}-C}
\STATE{Generate CIRs for communication ($h_c$) and sensing ($h_s$) via step \emph{4)} in Section \ref{section4}-C and (\ref{eqn_1}).}
\end{algorithmic}
\end{algorithm}

\subsection{Clustering Results of Measured MPCs}

Fig. \ref{fig_5a}-\ref{fig_5c} show the KPM-JCA results of measured MPCs for three ISAC channels, where the communication RX antenna is located at positions 1, 2, and 3 (as shown in Fig. \ref{fig_3}), respectively. The circle's center, angle, and radius meanings have the same meanings as in Fig. \ref{fig_4}, and different clusters are marked with different serial numbers and colors. As demonstrated in Fig. \ref{fig_5}, some clusters contain communication and sensing MPCs with similar attributes (i.e., angle and delay) at the same time, which are labeled by black numbers in these figures. For example, the clusters 1-4, 7, 10-11, and 15 in Fig. \ref{fig_5a}. Clusters 1-4 and 15 are contributed by the shared south wall (compared with Fig. \ref{fig_4a} and Fig. \ref{fig_4b}), cluster 7 corresponds to the shared north wall, and clusters 10-11 are associated with the east pillar and wall. These measured clusters validate the realistic shared clusters as expressed in model (\ref{eqn_1}). Similarly, clusters 1, 5-9, 11, 13, and 15 are shared clusters in Fig. \ref{fig_5b}. In Fig. \ref{fig_5c}, clusters 2-3, 5-7, and 10 are shared clusters. They are all contributed by the same shared scatterer and reflect the scatterer properties in terms of delay and angle. Moreover, the communication or sensing clusters that contain only communication or sensing MPCs, besides shared clusters, are annotated by the red solid and blue dashed circles in Fig. \ref{fig_5}.

As corroborated in Fig. \ref{fig_5}, the sensing clusters are more numerous and more discretely distributed, whereas communication clusters are fewer in number and have a more concentrated and sparse distribution. This observation is determined by the communication and sensing channel characteristics, and aligns with the channel scatterer distribution depicted in Fig. \ref{fig_4}. As previously stated, almost all scatterers in our measurement scenarios can be sensed, resulting in sensing clusters also covering information about the entire environment. However, in ISAC channels 1 and 2, as shown in Fig. \ref{fig_5a} and \ref{fig_5b}, cluster 14 only contains communication MPCs, which is at LOS condition. This is due to the communication RX not being located in the sensing measurement environment. In the case of NLOS condition, as shown in Fig. \ref{fig_5c}, the shared and communication cluster sets are equivalent. 

Due to this completeness of environment in our sensing measurements, $\text{SD}_s$ can be used as a benchmark to measure the degree of ISAC channel similarity. The values of $\text{SD}_s$ in three measured ISAC channels are 0.65, 0.64, and 0.54, respectively. These values are influenced by the realistic various channel environment and are higher at typical communication LOS condition (e.g., channels 1 and 2) compared to typical NLOS condition (e.g., channel 3). Specifically, the SD metric can be applied to guide the power configuration of the ISAC channels in modeling simulation. As the SD value increases, the correlation of ISAC channels grows, providing more opportunities to achieve the complementary auxiliary functions of communication and sensing in ISAC systems. More quantitative results of joint clustering are given in Table \ref{table_2}.

\subsection{Stochastic Characteristic Analysis}

\begin{figure}[!h]
\centering
\subfloat[]{\includegraphics[width=1.725in]{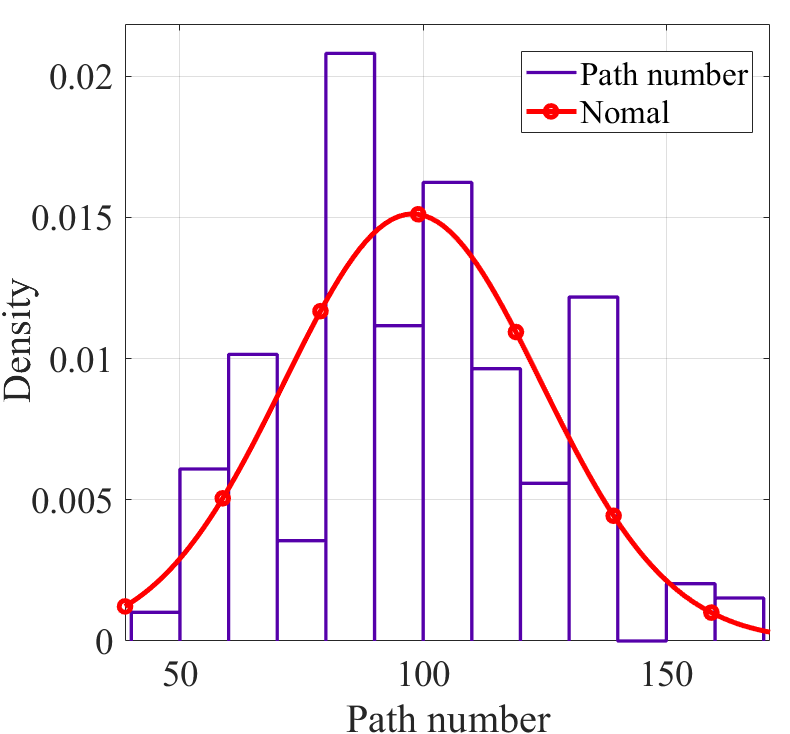}
\label{fig_7a}}
\subfloat[]{\includegraphics[width=1.725in]{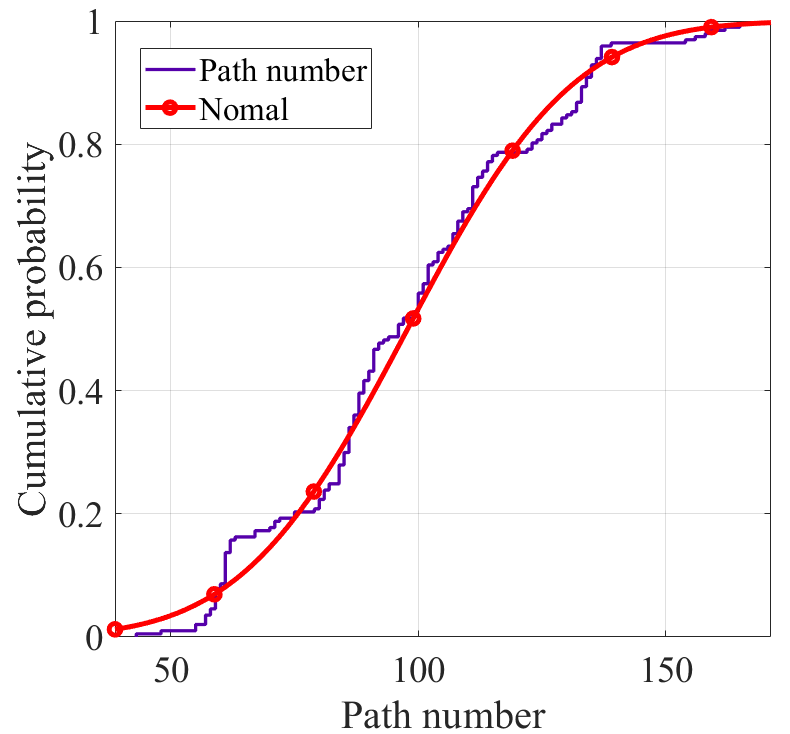}
\label{fig_6a}}\\
\subfloat[]{\includegraphics[width=1.725in]{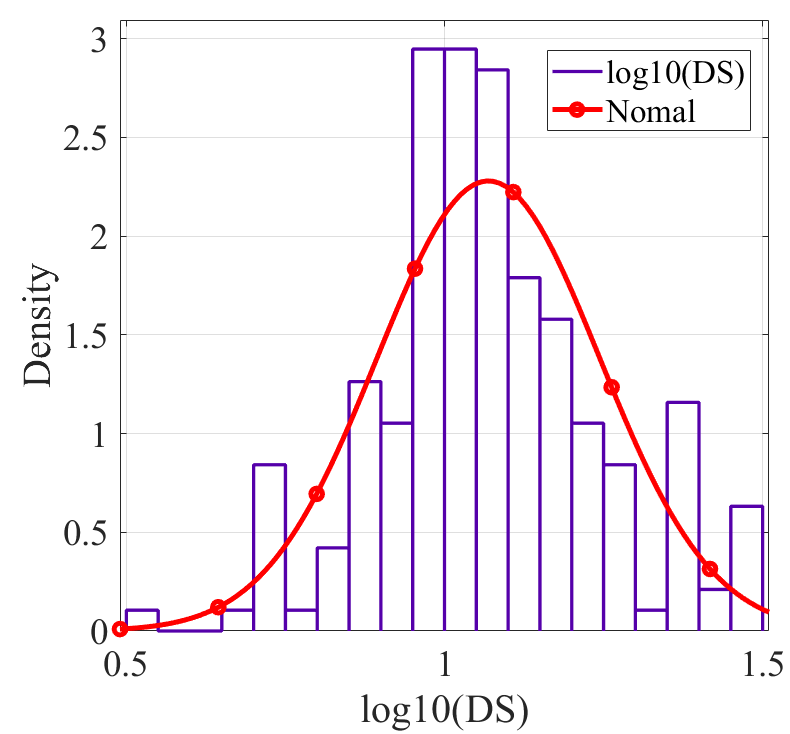}
\label{fig_7b}}
\subfloat[]{\includegraphics[width=1.725in]{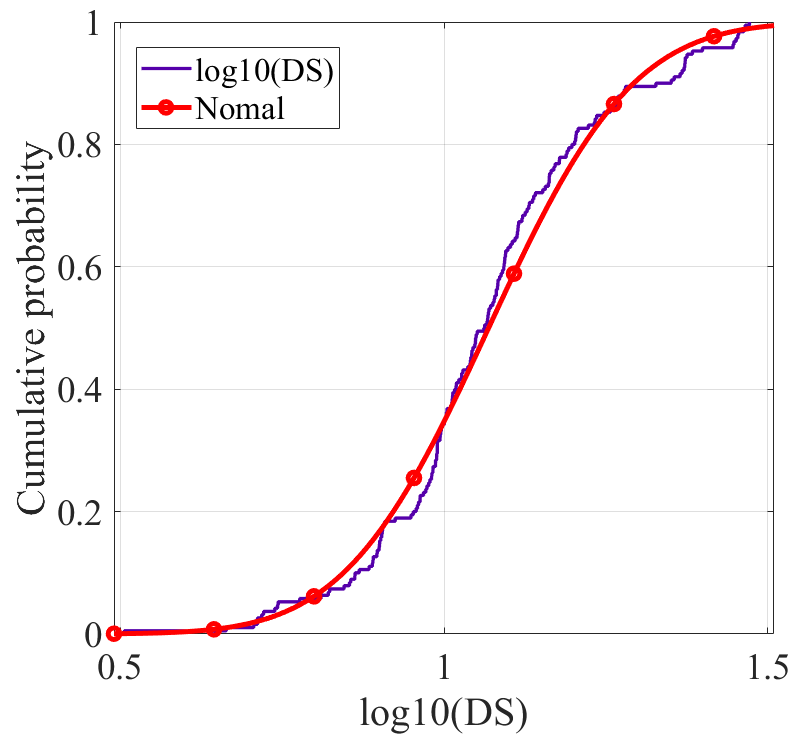}
\label{fig_6b}}\\
\subfloat[]{\includegraphics[width=1.725in]{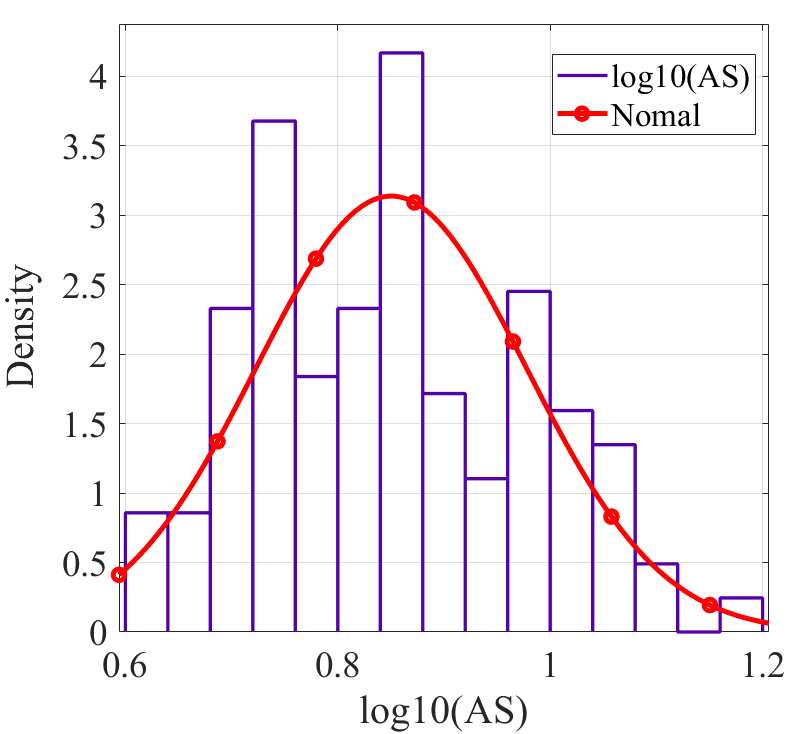}
\label{fig_7c}}
\subfloat[]{\includegraphics[width=1.725in]{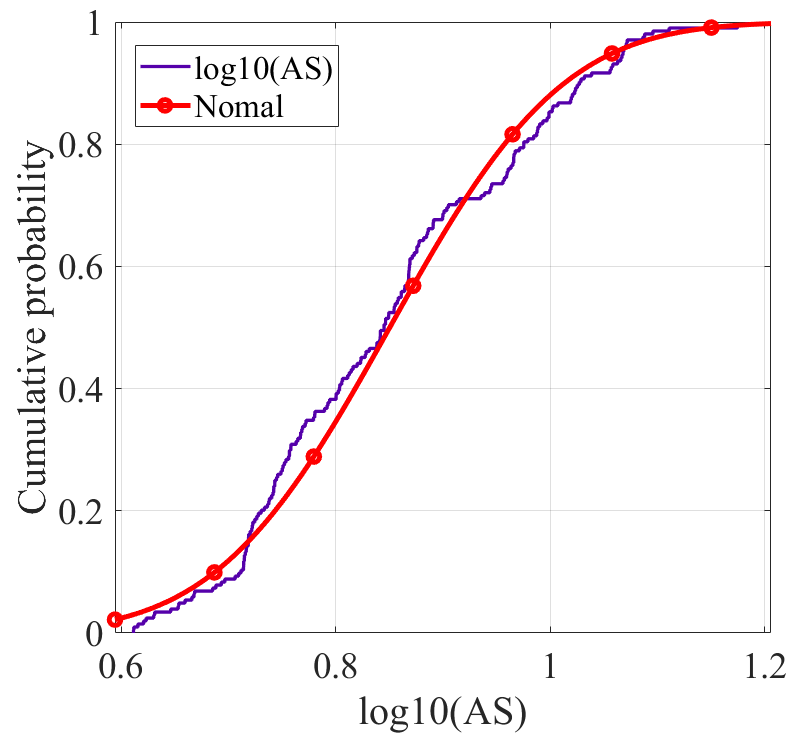}
\label{fig_6c}}
\caption{Measured PDFs and CDFs of shared intra-cluster parameters, including (a) (b) path number, (c) (d) DS, and (e) (f) AS.}
\label{fig_6}
\end{figure}

Based on the joint clustering results, we further acquire the stochastic intra- and inter- cluster parameters, including the path number, the root mean square DS (rms DS), and the rms AS on all snapshots of ISAC channel measurements. For example, the rms DS in $k^{th}$ cluster is calculated as

\begin{gather}
\label{equ_7}
\mu_{\tau,k}=\left(\sum\limits_{m=1}^{M_k}\tau_m\cdot p_m\right)/\sum\limits_{m=1}^{M_k}p_m,\\
\label{equ_8}
\sigma_{\tau,k}=\sqrt{\sum\limits_{m=1}^{M_k}(\tau_m-\mu_{\tau,k})^2\cdot p_m/\sum\limits_{m=1}^{M_k}p_m},
\end{gather}
where $\mu_{\tau,k}$ indicates the mean delay of the $k^{th}$ cluster. $\tau_m$ and $p_m$ are respectively the delay and normalized power of the $m^{th}$ MPC. $M_k$ is the total number of the MPCs in the $k^{th}$ cluster. Replacing $\tau$ with $\theta$ in the above formulas, the azimuthal AS can be calculated similarly. Besides, a method proposed in \cite{dong2007cluster} is utilized here to avoid angular ambiguity.

In terms of the intra-cluster parameters, this paper focuses primarily on the research of shared clusters, given that the independent communication or sensing clusters would follow the traditional rules. The Probability Density Function (PDF) and Cumulative Distribution Function (CDF) of intra-cluster path numbers, $\text{log}_{10}(\text{DS})$, and $\text{log}_{10}(\text{AS})$ are drawn in Fig. \ref{fig_6}. These distributions are fitted through the normal distribution $\mathcal{N}(\mu,\sigma^2)$, which is consistent with ITU-R M.2412 \cite{itu2412}. The mean values (i.e., $\mu$) of the distribution are 98, 1.07 (11.75 ns), and 0.85 (7.08$^\circ$), respectively. The greater the variance (i.e., $\sigma$), the more dynamic the parameter value. Moreover, the stochastic parameters of shared sub-clusters (marked with braces in model (\ref{eqn_1})) are further extracted and listed in Table \ref{table_3}. 

By applying all MPC parameters from a single clustering result into formulas (\ref{equ_7}) and (\ref{equ_8}), the inter-cluster parameters can be calculated. The cluster distribution of the delay and angular domain becomes more discrete as the value of the inter-cluster parameters (rms DS and AS) increases. The DS and AS averages of joint, communication, and sensing clusters (i.e. $N$, $N_c$, and $N_s$ in model (\ref{eqn_1})) are shown in Table \ref{table_4}. Based on the obtained quantitative results, the stochastic characteristics exhibited by the clusters in the delay and angular dimensions are analyzed below.

\begin{table}[t]
\caption{Stochastic Parameters within Shared Clusters and Sub-Clusters\label{table_3}}
\centering
\begin{tabular}{cccc}
\hline
\multirow{3}{*}{\makecell[c]{Parameter}} & \multirow{3}{*}{\makecell[c]{Shared clusters\\($\mu$,$\sigma$)}} & \multicolumn{2}{c}{Shared sub-clusters}\\
\cline{3-4}
 & & \makecell[c]{Communication\\($\mu$,$\sigma$)} & \makecell[c]{Sensing\\($\mu$,$\sigma$)}\\ 
\hline
\makecell[c]{Path number} & \makecell[c]{(98,695)} & \makecell[c]{(24,153)} & \makecell[c]{(75,867)}\\
\makecell[c]{$\text{log}_{10}(\text{DS})$} & \makecell[c]{(1.07,0.03)}  & \makecell[c]{(0.55,0.19)} & \makecell[c]{(1.08,0.06)}\\
\makecell[c]{$\text{log}_{10}(\text{AS})$} & \makecell[c]{(0.85,0.02)} & \makecell[c]{(0.73,0.03)} & \makecell[c]{(0.88,0.04)}\\
\hline
\end{tabular}
\end{table}

\begin{table}[t]
\caption{Main Inter-Cluster Parameters
}\label{table_4}
\centering
\begin{tabular}{ccc}
\hline
Cluster categories &  rms DS (ns) & rms AS ($^\circ$)\\
\hline
Total joint clusters & 24.77 & 62.35\\
Communication clusters & 13.87 & 38.66\\
Sensing clusters & 28.82 & 91.29\\
\hline
\end{tabular}
\end{table}

\emph{1) Delay characteristics:} 
As can be seen from the main inter-cluster parameters in Table \ref{table_4}, the rms DS value of sensing clusters (28.82 ns) is rough twice the corresponding value of communication clusters (13.87 ns), and the value of joint clusters (24.77 ns) reflects the average effect of communication and sensing. These numerical results demonstrate the discrete properties of sensing clusters as well as the centrality and sparsity of communication clusters in the delay domain.

The stochastic parameters within shared clusters and sub-clusters in Table \ref{table_3} demonstrate that the $\mu$ value of the shared sensing sub-clusters in DS fitted distribution (1.08) is higher, while the $\sigma$ value (0.06) is lower than that of shared communication sub-clusters (0.55 and 0.19). This finding corroborates that the MPCs within shared sensing sub-clusters have a more discrete distribution but a more stable structure than that of communication in the delay domain. Furthermore, due to the larger path number, the shared sensing sub-clusters have a greater impact on the overall, and their parameter values are closer to those of the shared clusters (1.07 and 0.03).

\emph{2) Angular characteristics:} 
As shown in Table \ref{table_4}, the mean inter-cluster AS values of communication and sensing clusters (38.66$^\circ$ and 91.29$^\circ$) prove their centralized and discrete distribution in the angular domain, respectively. The value of sensing clusters is approximately 2.4 times greater than that of communication clusters, and they balance the value of the joint clusters (62.35$^\circ$).

The results in Table \ref{table_3} show that the $\mu$ and $\sigma$ values of the shared sensing sub-clusters in AS fitted distribution (0.88 and 0.04) are higher than those of communication sub-clusters (0.73 and 0.03). It corroborates that, in the angular domain, the MPC distribution within shared communication sub-clusters is more concentrated and sparse compared to sensing clusters, and it exhibits a more stable structure. Moreover, the $\sigma$ of all $\text{log}_{10}(\text{AS})$ distributions in Table \ref{table_3} are relatively low and close to each other, and the shared sensing sub-clusters have a slightly greater impact on the overall $\mu$ value of shared clusters (0.85). 

Compared the rms AS values for all types of clusters to the DS values, the inter-cluster AS values are relatively larger, while the intra-cluster AS values are relatively smaller. This indicates that the measured ISAC clusters demonstrate better aggregation characteristics in the angle domain.

\subsection{Simulation Validation}
This paper develops and performs preliminary simulations based on the measurement results to validate the practicality and effectiveness of the proposed model. The Indoor Hotspot (InH) scenario from ITU-R M.2412 \cite{itu2412}, which is adapted to the measurement environment of this paper, is considered as the baseline scenario. The simulation configuration consists of a single base station and user terminal, both equipped with a single antenna. The center frequency is set as 28 GHz, and the system bandwidth is set as 1 GHz, consistent with our measurement campaign.

In practical ISAC systems, the channel sharing feature mainly manifests in the angular similarity of shared clusters, which has been observed from the measurements discussed earlier. Therefore, this paper focuses on channel realization at the small-scale level, including the generation of delays, powers, and angles, of the ITU-R M.2412 standards. Based on the same set of environment layouts, the ISAC model implementation steps of small-scale parameters are summarized as follows.

\begin{figure*}[t]
\centering
\subfloat[]{\includegraphics[width=1.735in]{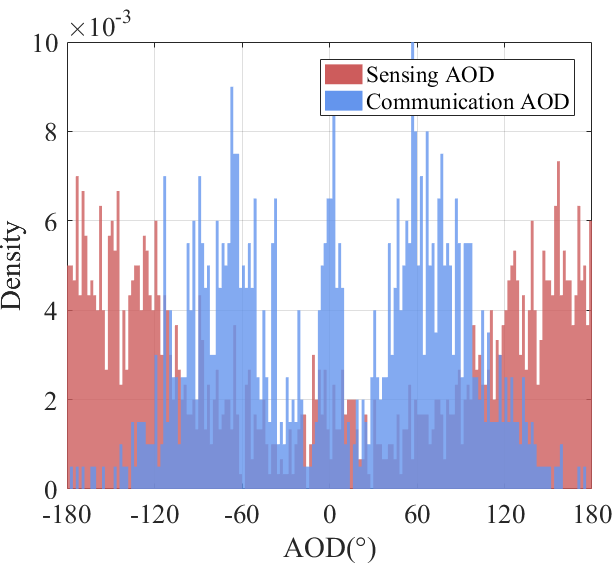}
\label{fig_12a}}
\hfil
\subfloat[]{\includegraphics[width=1.735in]{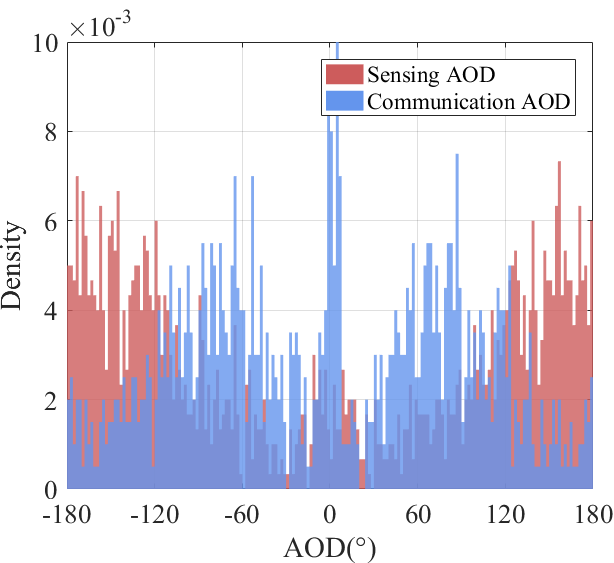}
\label{fig_12b}}
\subfloat[]{\includegraphics[width=1.735in]{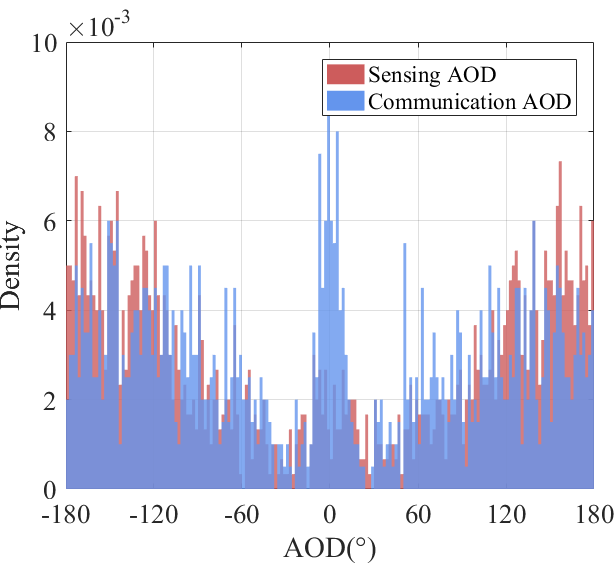}
\label{fig_12c}}
\subfloat[]{\includegraphics[width=1.735in]{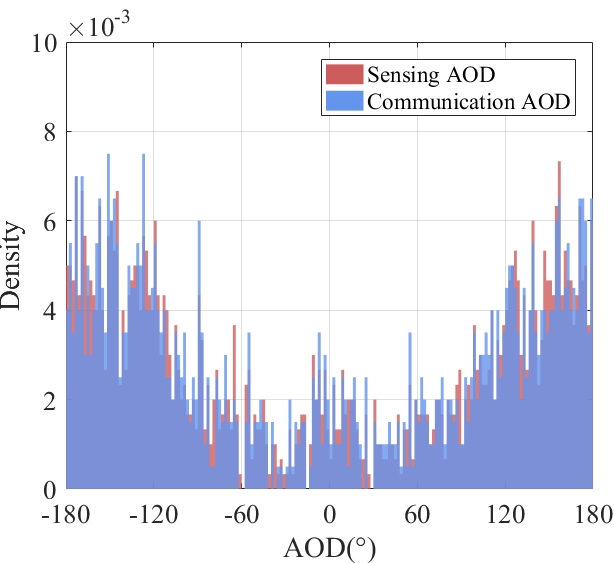}
\label{fig_12d}}
\caption{Simulation PDFs of cluster centroid AODs, when the shared cluster number is (a) 0, (b) 2, (c) 6, and (d) 10. The total cluster number of communication and sensing are $N_c=10$ and $N_s=15$, respectively. The sensing PDF (red area) is repeated as a benchmark in the figures. As the $N_0$ proportion increases (from (a) to (d)), the communication PDF (blue area) gradually converges with the red one.}
\label{fig_12}
\end{figure*}

\emph{1) Generate parameters of sensing clusters.} The measured clustering results of sensing channels (such as inter-cluster DS and AS values in Table \ref{table_4}) are applied in simulation to generate stochastic parameters of sensing clusters with the number $N_s$. The other configurations adopt the default values of the standards.

\emph{2) Generate shared parameters of communication clusters.} $N_0$ clusters are randomly selected from the sensing clusters generated in the previous step as shared sub-clusters, whose cluster centroid angles are reused as the values of the shared communication sub-clusters. The other parameters of shared communication sub-clusters continue to be generated stochastically based on the standards and measured data.

\emph{3) Generate stochastic parameters of communication clusters.} Similarly to step \emph{1)}, the measured clustering results (DS and AS values) are applied to generate stochastic parameters of non-shared communication clusters with the number $N_1=N_c-N_0$. The other configurations adopt the default values of the standards.

\emph{4) Generate ISAC channel response.} The global ISAC channel response is generated by superposing channel responses of all the above clusters. 

\begin{figure}[h]
\centering
\includegraphics[width=3.4in]{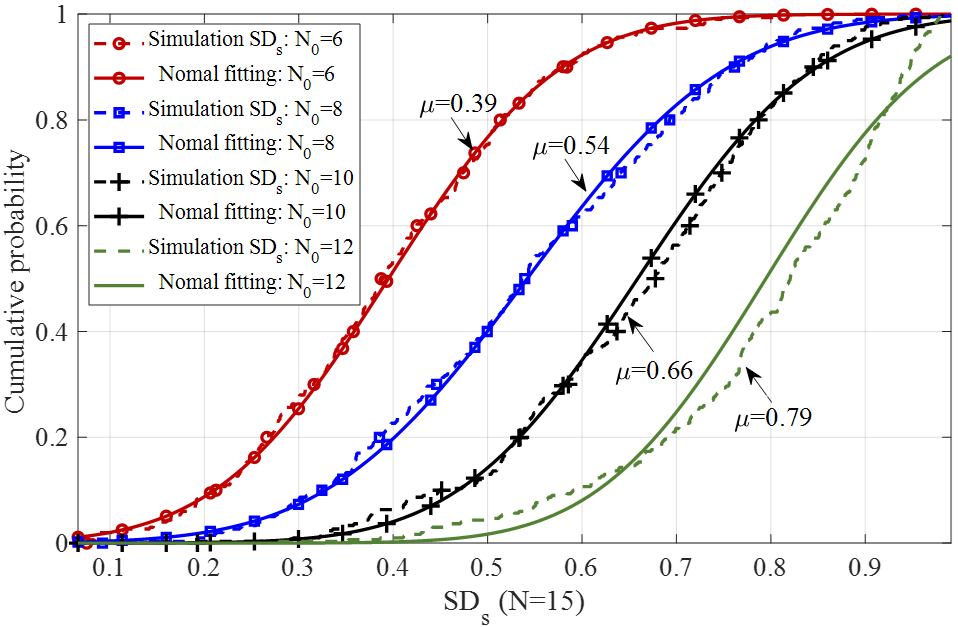}
\caption{Simulation CDFs of $\text{SD}_s$ with varying numbers of shared clusters, $N_0$, when the total cluster number is $N=N_s=15$. The value of the $\text{SD}_s$ increases as the proportion of $N_0$ increases.}
\label{fig_11}
\end{figure}

Firstly, we set the total cluster number $N=15$, and disregard the generation of non-shared communication clusters in simulation step \emph{3)} (i.e., $N_s=N$ and $N_c=N_0$), which is adapted to the measurements in this paper. Fig. \ref{fig_11} presents the CDFs of $\text{SD}_s$ when the number of shared clusters $N_0$ (i.e., communication clusters $N_c$) is 6, 8, 10, and 12. The dashed lines depict 300 sets of simulation results, and the solid lines are the fitting curves of the normal distribution. These CDFs corroborate that a larger proportion of $N_0$ leads to a higher value of $\text{SD}_s$. The desired $\text{SD}_s$ value can be obtained by controlling the appropriate $N_0$. When $N_0=10$, the mean value of $\text{SD}_s$ ($\mu=0.66$) is close to the results of measured ISAC channels 1 (0.65) and 2 (0.64), as listed in Table \ref{table_2}. As for ISAC channel 3, the proportion of $N_0$ is $N_0/N=6/11=54.5\%$, whose realistic $\text{SD}_s$ value is the same as curve $\mu=0.54$ ($N_0=8$). The $N_0$ proportion of this curve is $N_0/N=8/15=53.3\%$, which is close to the proportion of ISAC channel 3.

The non-shared communication clusters $N_1$ should be considered in the simulation, although they are not intuitively represented in the measurement results of this paper. In Section \ref{section3}, the possible situations for different types of scatterers and clusters have been analyzed. Since the shared sub-clusters are directly associated with the cluster centroid angles in the simulation configuration, we examine the effect of changing the shared cluster settings on the ISAC channels using the cluster centroid AOD as an example. Here, we set the total cluster number of communication and sensing as $N_c=10$ and $N_s=15$, respectively. The LOS angle is set as zero-degree. We perform 100 sets of simulations by applying the inter-cluster AS of stochastic sensing and communication clusters as 91.29$^\circ$ and 38.66$^\circ$, respectively, based on the measured clustering results in Table \ref{table_3}. For each simulation, the stochastic parameters of sensing clusters $N_s$ are generated according to the simulation step \emph{1)}. Then, communication parameters are simulated when the shared cluster number, $N_0$, is 0, 2, 6, and 10, following the steps \emph{2)} and \emph{3)}. 

The PDFs of communication and sensing AODs are shown in Fig. \ref{fig_12}, where the sensing PDF (indicated by the red area) is repeated as a benchmark in these figures. Fig. \ref{fig_12a} shows the random generation result of communication and sensing AODs with their respective AS values. (Note that the Power Angular Spectrum (PAS) in the azimuth satisfies the wrapped Gaussian distribution, which is consistent with the standards. However, this paper ignores the power dimension to simplify and focus on the angular varies with the different number of shared clusters.) As the proportion of $N_0$ increases (from Fig. \ref{fig_12a} to Fig. \ref{fig_12d}), the PDF of communication AODs (indicated by the blue area) gradually converges with the sensing PDF, and channels will exhibit more similar characteristics in the angular domain.

\section{Conclusion}\label{section5}
This paper proposes a shared cluster-based stochastic channel model for ISAC systems and validates it according to realistic channel measurements and simulations. Firstly, we conduct the channel measurement campaign in typical LOS and NLOS indoor scenarios at 28 GHz and obtain the PADPs. The scatterers shared by communication and sensing channels are intuitively observed. Based on the concept of shared clusters (contributed by the shared scatterers), a stochastic channel model expressed by the superposition of shared and non-shared clusters is proposed to capture the sharing feature. The SD metric is also introduced to define the power ratio of the shared clusters in the channels. Correspondingly, the KPM-JCA is novelly introduced, which enables extracting the shared and non-shared clusters for ISAC channels. Finally, based on the proposed model and clustering algorithms, stochastic inter- and intra- cluster parameters, including the DS, AS, and SD, are acquired from the measurement results. And the experimental validation of the proposed model with parameterization is performed. The results demonstrate that the communication clusters exhibit a more concentrated and sparse distribution while sensing clusters are distributed more discretely. In addition, the shared MPCs show stronger clustering features in the angular domain than in the delay domain. The clustering results of measured MPCs confirm and quantify the realistic characteristics of shared clusters in ISAC channels, where the SD is calculated as 0.65, 0.64, and 0.54 in the two LOS and one NLOS measurements, respectively. The simulation work further validates that the channel SD increases as the shared cluster number increases of the proposed model, which is applicable to the development of ISAC systems.

For future work, conducting additional measurements and practical models is necessary to conduct comprehensive research on ISAC channels, despite the potential increase in overhead costs. It would be valuable to arrange a double-directional channel measurement campaign to analyze angular characteristics at both TX and RX sides. Besides, the ISAC channel measurements at more frequency bands, e.g., sub-6 GHz, terahertz (THz), and in more typical scenarios, e.g., outdoor, Industrial Internet Of Things (IIoT), are worth being conducted to discover some interesting regularities of sharing feature, which help us further develop the ISAC channel model. Furthermore, it will also be beneficial to establish a standardized ISAC channel modeling framework with more specific statements and details.

\bibliography{reference}

 
\vspace{11pt}
\begin{IEEEbiography}[{\includegraphics[width=1in,height=1.25in,clip,keepaspectratio]{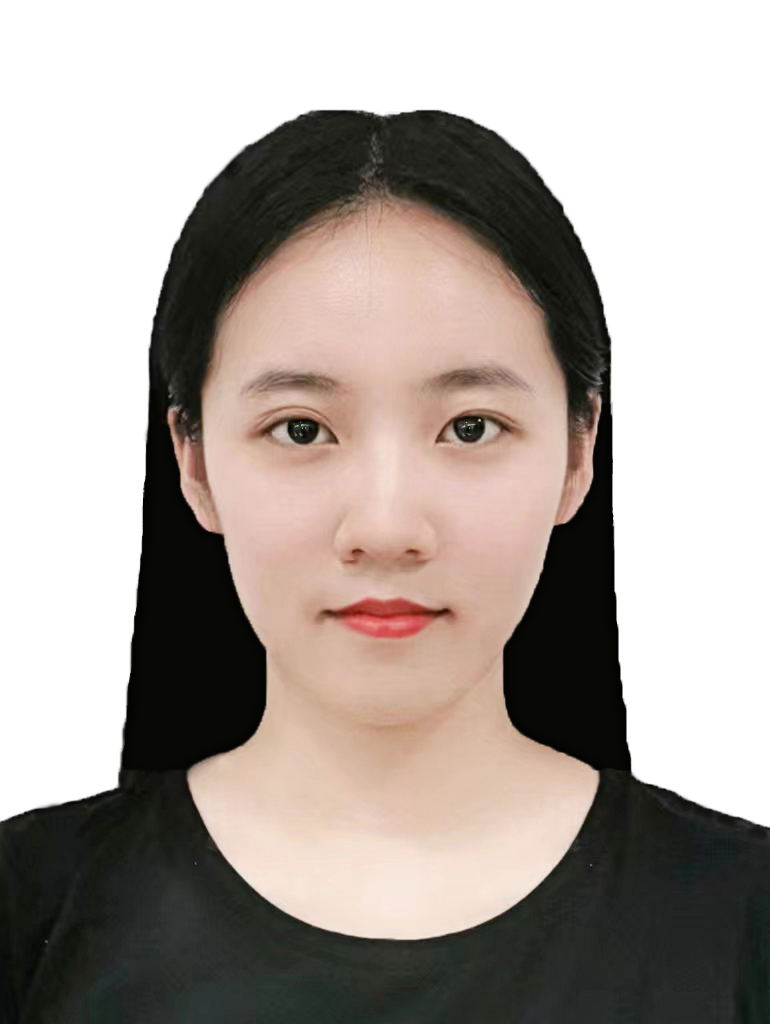}}]{Yameng Liu} (Graduate Student Member, IEEE)
received the B.S. degree in communication engineering from Minzu University of China in 2021. She is currently pursuing the
Ph.D. degree with the State Key Laboratory of Networking and Switching Technology, Beijing University of Posts and Telecommunications (BUPT). Her current research interests include channel measurements and modeling, integrated sensing and communication,  and millimeter wave.
\end{IEEEbiography}

\begin{IEEEbiography}[{\includegraphics[width=1in,height=1.25in,clip,keepaspectratio]{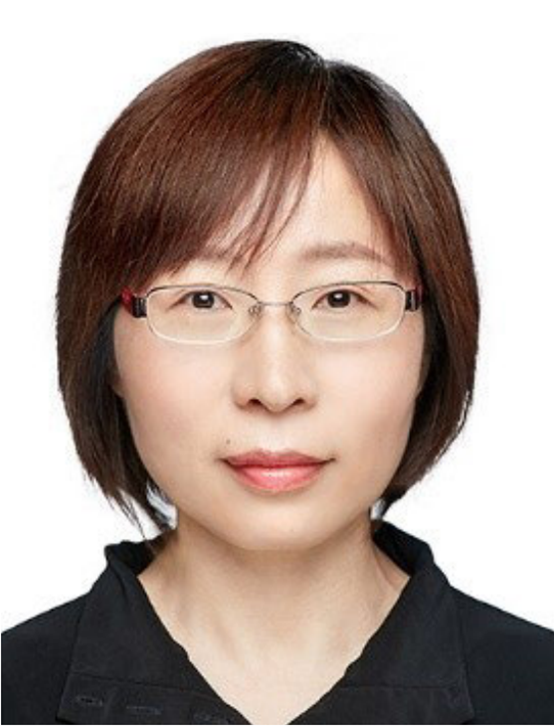}}]
{Jianhua Zhang} (Senior Member, IEEE) received the Ph.D. degree from the Beijing University of Posts and Telecommunications (BUPT) in 2003. She is currently a Professor with BUPT, the China Institute of Communications Fellow, and the Director of the BUPT--CMCC Joint Research Center. She has published more than 200 papers and authorized 40 patents. She received several paper awards, including 2019 SCIENCE China Information Hot Paper, 2016 China Comms Best Paper, and 2008 JCN Best Paper. She received several prizes for her contribution to ITU--R 4G channel model (ITU--R M.2135), 3GPP relay channel model (3GPP 36.814), and 3GPP 3D channel model (3GPP 36.873). She was also a member of 3GPP  ``5G channel model for bands up to 100 GHz''. From 2016 to 2017, she was the Drafting Group (DG) Chairperson of ITU--R IMT--2020 Channel Model and led the drafting of the ITU--R M. 2412 Channel Model Section. She is also the Chairwomen of the China IMT--2030 Tech Group--Channel Measurement and Modeling Subgroup and works on 6G channel model. Her current research interests include beyond 5G and 6G, artificial intelligence, data mining, channel modeling for integrated sensing and communication, massive MIMO, millimeter wave, THz, visible light channel modeling, channel emulator, and OTA test.
\end{IEEEbiography}

\begin{IEEEbiography}[{\includegraphics[width=1in,height=1.25in,clip,keepaspectratio]{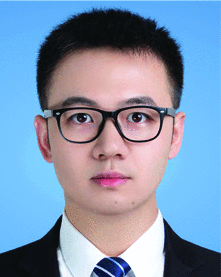}}]
{Yuxiang Zhang} (Member, IEEE) is the associate researcher in the State Key Laboratory of Networking and Switching Technology, BUPT, China. He received the B.S. degree in electronic information engineering from the Dalian University of Technology in 2014 and the Ph.D. degree from the BUPT in 2020. From 2018 to 2019, he was a Visiting Scholar with the University of Waterloo. He has authored and co-authored more than 30 papers in refereed journals and conference proceedings. His current research interests include massive/holographic MIMO, integrated sensing and communication, and reconfigurable intelligent surface channel measurement and modeling.
\end{IEEEbiography}

\begin{IEEEbiography}[{\includegraphics[width=1in,height=1.25in,clip,keepaspectratio]{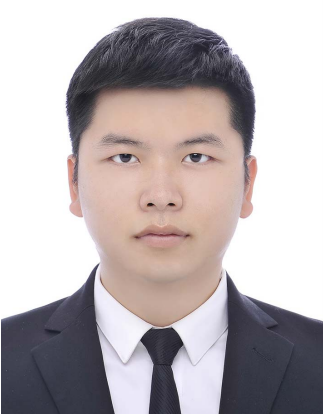}}]
{Zhiqiang Yuan} (Graduate Student Member, IEEE) received the B.S. degree from the Beijing University of Posts and Telecommunications (BUPT), Beijing, China, in 2018, where he is currently pursuing the Ph.D. degree. He had been a Visiting Ph.D. Student with the Antennas, Propagation, and Millimeter-Wave Systems (APMS) Section, Aalborg University, Aalborg, Denmark, from Jul. 2021 to Jul. 2023. He currently focuses on radio channel sounding and modeling for massive MIMO, millimeter-wave (mmWave), and THz systems, antenna array signal processing, and channel parameter estimation.
\end{IEEEbiography}

\begin{IEEEbiography}[{\includegraphics[width=1in,height=1.25in,clip,keepaspectratio]{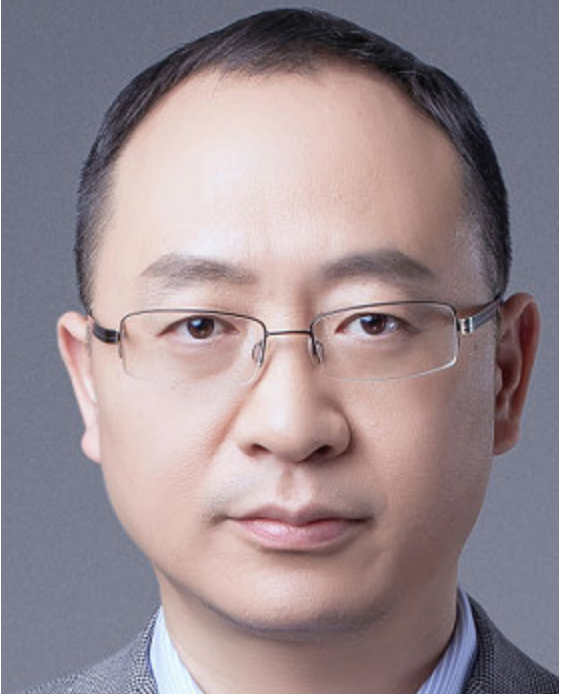}}]
{Guangyi Liu} (Member, IEEE) received the Ph.D. degree from the Beijing University of Posts and Telecommunications, Beijing, China, in 2006. He is currently a Chief Scientist with 6G in China Mobile Communication Corporation (CMCC), the Co--Chair of the 6G Alliance of Network AI (6GANA), the Vice Chair of THz Industry Alliance in China, and the Vice Chair of the Wireless Technology Working Group of IMT--2030 (6G) Promotion Group supported by the Ministry of Information and Industry Technology of China. He has been leading the 6G Research and Development of CMCC since 2018. He has acted as the Spectrum Working Group Chair and the Project Coordinator of LTE Evolution and 5G eMBB in Global TD--LTE Initiative (GTI) (2013--2020) and led the industrialization and globalization of TD--LTE evolution and 5G eMBB.
\end{IEEEbiography}

\vfill

\end{document}